\documentclass[useAMS,usenatbib,epsfig]{mn2e}
\usepackage{graphicx,stfloats,pdflscape,here}

\title[]{Sejong Open Cluster Survey (SOS). \\
III. The Young Open Cluster NGC 1893 in the H {\Large \textsc{II}} Region W8}
\author[B. Lim et al.]
  {Beomdu Lim,$^{1,2}$\thanks{Corresponding author, Korean Research Council of Fundamental Science \& Technology Research Fellow, E-mail:bdlim1210@kasi.re.kr}
  Hwankyung Sung,$^2$\thanks{E-mail:sungh@sejong.ac.kr} 
  Jinyoung S. Kim,$^3$  
  Michael S. Bessell,$^4$ 
 \newauthor and Byeong-Gon Park$^1$ \\
  $^1$ Korea Astronomy and Space Science Institute, 776 Daedeokdae-ro, Yuseong-gu, Daejeon 305-348, Korea \\
  $^2$ Department of Astronomy and Space Science, Sejong University, 209 Neungdong-Ro, Gwangjin-gu, Seoul 143-747, Korea\\
  $^3$ Steward Observatory, University of Arizona, 933 N. Cherry Ave. Tucson, AZ 85721-0065, USA\\
  $^4$ Research School of Astronomy and Astrophysics, Australian National University, MSO, Cotter Road, Weston, ACT 2611, Australia\\
  }
\date{Released 2013 Xxxxx XX}

\pagerange{\pageref{firstpage}--\pageref{lastpage}} \pubyear{2013}

\def\LaTeX{L\kern-.36em\raise.3ex\hbox{a}\kern-.15em
    T\kern-.1667em\lower.7ex\hbox{E}\kern-.125emX}

\begin{document}

\label{firstpage}

\maketitle

\begin{abstract}
We present a $UBVI$ and H$\alpha$ photometric study of the young open cluster NGC 1893 in the 
H {\scriptsize \textsc{II}} region W8 (IC 410 or Sh 2-236). A total of  65 early-type members are selected from photometric 
diagrams. A mean reddening of the stars is $\langle E(B-V) \rangle = 0.563 \pm 0.083$ mag. The published 
photometric data in the near- and mid-infrared passbands are used to test the reddening law 
toward the cluster, and we confirm that the reddening law is normal ($R_V = 3.1$). Zero-age main sequence 
fitting gives a distance modulus of $V_0 - M_V = 12.7 \pm 0.2$ mag, equivalent 
to $3.5 \pm 0.3$ kpc. From H$\alpha$ photometry 125 H$\alpha$ emission stars and 
candidates are identified as pre-main sequence (PMS). The lists of young stellar 
objects and X-ray sources published by previous studies allow us to select a large number of PMS members down 
to $1M_{\sun}$. Isochrone fitting in the Hertzsprung-Russell diagram gives a turn-off age of 
1.5 Myr and the median age of 1.9 Myr from the PMS members with a spread of $\sim$ 5 Myr. 
We derive the initial mass function (IMF) for stars with mass larger than $1M_{\sun}$. The slope of the 
IMF ($\Gamma = -1.3 \pm 0.1$) is well consistent with the Salpeter/Kroupa IMF. A total mass of 
the cluster appears to be in excess of 1,300 $M_{\sun}$. Finally, we estimate the mass 
accretion rate of 82 PMS members in the mass range of $0.6M_{\sun}$ to $5M_{\sun}$.  
\end{abstract}


\section{Introduction}
NGC 1893 is the young open cluster in the H {\scriptsize \textsc{II}} region W8 (IC 410 or Sh 2-236) which 
is a part of the Auriga OB2 association. The cluster incubates 5 O-type stars, HDE 242935 [O6.5V((f))z], 
BD +33 1025 [O7V(n)z], HDE 242908 [O4.5V(n)((fc))], HDE 242926 [O7Vz], and TYC 2394-1214-1 
\citep{HA65,SAW11}, as well as two prominent emission nebulae, Sim 129 and 130. These O-type stars 
are thought to be the main ionizing sources within the region. The morphology of the emission 
nebulae indicates that star formation is currently taking place. From an analysis of two stars in the cluster, 
\citet{DC04} found $\sim 0.26$ dex lower abundance of light elements than that found in solar neighbourhood clusters. 
If this is correct, it implies that the stars in NGC 1893 are forming in a low-metallicity environment. It is considered
that  star-forming activities in the outer Galaxy may be less vigorous as a result of several contributing factors - a low-metallicity
environment, weak interactions with spiral arms, and a lack of supernova explosions. Thus, NGC 1893 could be a very interesting
laboratory for studying star formation processes and the initial mass function (IMF) in a different environment 
from the solar neighbourhood (\citealt{CMP08} and references therein).  

Many photometric survey studies \citep{JHIMH61,B63,HA65,M72,MJD95,LGM01} involving NGC 1893 have 
been conducted in the optical passbands. These studies presented useful photometric data and 
fundamental parameters, such as reddening, distance, and age of the cluster. However, the photometric 
data are not deep enough to study more details for pre-main sequence stars (PMS). Since \citet{VRCG99} and 
\citet{MBN01} predicted the presence of a large number of PMS stars in the cluster, many researchers 
became interested in the PMS population. Several spectroscopic observations have confirmed that 
the majority of PMS stars with emission lines are found in the cluster core as well as in the vicinity of Sim 129 and 130 
\citep{MN02,MSBPB07,SPO07,NMIB07}. In addition, many PMS stars, which exhibit 
near-infrared (NIR) excess from their circumstellar discs, were identified using the Two Micron 
All Sky Survey (2MASS; \citealt{2mass}) data. From the age distribution of the PMS stars inferred from the 
($V, V-I$) colour-magnitude diagram (CMD) \citet{MSBPB07} and \citet{SPO07} argued that star 
formation has progressively taken place from the cluster centre toward the two emission nebulae. \citet{NMIB07} 
also described how the observed properties, such as the morphology of the nebulae and presence 
of young emission-line PMS stars away from the cluster, are very similar to the typical 
characteristics of triggered star formation delineated by \citet{W02}. However, the membership seemed 
to be limited to a small number of PMS stars with either H$\alpha$ emission or NIR excess emission. 

More detailed studies of the PMS population were made possible by several extensive observational programs. 
\citet{CMP08} presented a study of a large PMS population based on 
{\it Spitzer} mid-infrared (MIR) and {\it Chandra} X-ray data. They identified a number of PMS stars (7 Class 0/I, 
242 Class II, and 110 Class III candidates) and estimated the fraction of PMS stars with 
a circumstellar disc (67 per cent). Later, \citet{PSM11} made deep optical and NIR observations for NGC 1893. 
By defining a Q index from various colours, a total of 1034 Class II objects were newly identified. 
The authors found 442 PMS stars without a circumstellar disc from their X-ray catalogue, thereby 
estimating a disc fraction of 71 per cent, which is higher than that obtained by \citet{CMP08}. \citet{CMP12} investigated 
the coronal properties of the PMS stars based on their X-ray luminosities derived from spectral 
fitting and quantile analysis. They found that the X-ray luminosity of Class III objects appears to be 
higher than that of Class II objects at the same bolometric luminosity, and 
suggest it may reflect the disc locking in Class II objects. On the 
other hand, Class II objects exhibit higher variability and more frequent flares than those of Class III 
objects. Comparing the X-ray properties of stars in NGC 1893 with those of stars in Orion nebula cluster (ONC) the 
authors concluded that the coronal properties of stars formed in the outer Galaxy may be the same as those 
of stars in nearby star-forming regions. 

\citet{LPCMC12} identified 53 variable PMS stars through time series 
observations in the $V$ and $I$ bands. They found that the rotational period of PMS stars 
decreases with stellar age and mass, and that the amplitude in the light curves also declines 
with the same physical quantities. The former is compatible with the disc locking models. 
\citet{PSC13} compared the age of classical T-Tauri stars (CTTSs) with that of 
weak-line T-Tauri stars (WTTSs) based on a cumulative age distribution. Their result indicates 
that CTTSs and WTTSs are coeval and have similar properties.

The previously determined distance to NGC 1893 showed a broad range from 3.2 kpc to 6.0 kpc: 
3.2 -- 3.3 kpc \citep{H78,SPO07}; 3.6 kpc \citep{B63,C73b,LGM01,PSM11}; 
4.0 kpc \citep{JHIMH61,HA65,WH68,M72}; 4.4 kpc \citep{TCER91,MJD95}; 4.8 kpc \citep{F93}; 
6.0 kpc \citep{MBN01}. All the previous studies relied on photometric methods, such as 
zero-age main sequence (ZAMS) fitting, spectroscopic parallax, isochrone fitting, and an
H$\gamma$ -- $M_V$ relation. Although the normal reddening law ($R_V = 3.0$ -- 3.1) 
was adopted in these studies, the discrepancy in distances derived from different 
authors is not negligible. Since distance is the most important fundamental parameter 
in converting observational parameters to reliable absolute physical quantities, 
it is necessary to revisit its determination.

\begin{figure*}
\includegraphics[height=0.45\textwidth]{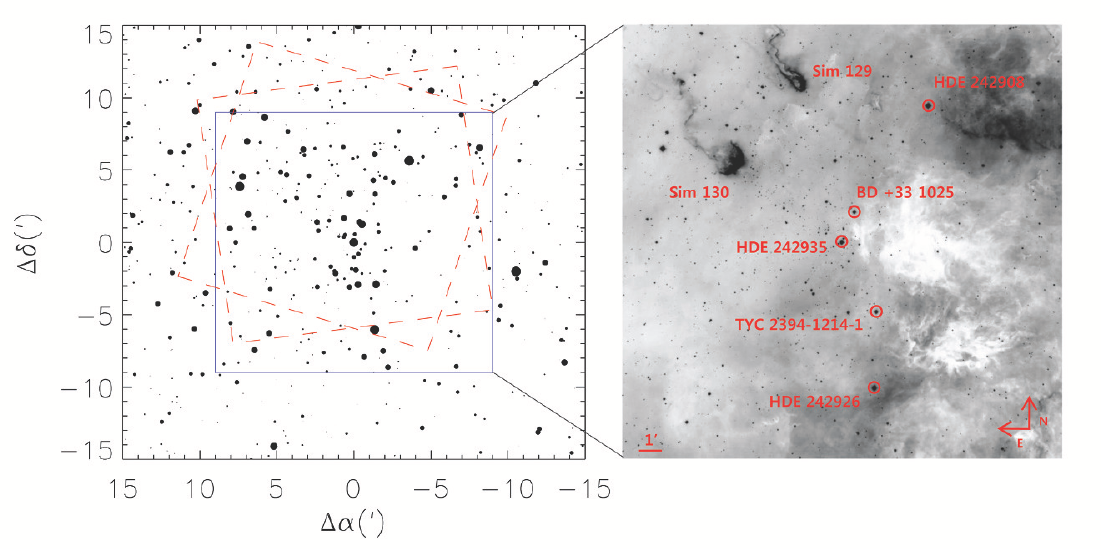}
\caption{Finder chart (left) and an H$\alpha$ image (right) for NGC 1893. The size of the circles in the finder chart is 
proportional to the brightness of the stars. The position of stars is relative to 
the O-type star HDE 242935 ($\alpha = 05^h \ 22^m \ 46.^{s}5$, $\delta = +33\degr \ 25' \ 11''$, 
J2000). A square (solid line) outlines the observed region in this work. Another two boxes (dashed line) 
display the {\it Chandra} ACIS fields of view \citep{CMP08}. The right-hand side panel shows the 
position of 5 O-type stars and two emission nebulae.  }
\label{fig1}
\end{figure*}

\begin{table*}
\begin{minipage}{150mm}
\caption{Photometric data and spectral type for 24 bright stars}
\begin{tabular}{lccclll}
\hline
ID & $V$ & $B-V$ & $U-B$ & reference & Spectral type & reference \\
\hline
BD +33 1025          & 10.31 & 0.26  & -0.73 & \citet{HJI61}   & O7V(n)z         & \citet{SAW11} \\
BD +33 1028          &   9.15 & 1.21  &  0.95  & \citet{HJI61}   &                       & \\
HDE 242855           & 10.65 & 0.06  & -0.35 & \citet{C73a}     &                        & \\
HDE 242908           &   9.05 & 0.27  & -0.72 & \citet{JM55}    &  O4.5V(n)((fc)) & \citet{SAW11}\\
HDE 242926           &   9.35 & 0.32  & -0.66 & \citet{JM55}    & O7Vz            & \citet{SAW11}\\
HDE 242935           &   9.43 & 0.20  & -0.73 & \citet{H56}     & O6.5V((f))z     & \citet{SAW11}\\
HDE 243018           & 10.94 & 0.21  & -0.77 & \citet{HJI61}    & B0V                & \citet{NMIB07}\\
TYC 2394-162-1    & 10.79 & 0.25  &  0.14 & \citet{HJI61}    &                      & \\
TYC 2394-450-1    & 11.47 & 0.17  & -0.66 & \citet{MJD95}  & B1III              & \citet{MJD95}\\
TYC 2394-503-1    & 11.08 & 0.58  &   0.03 & \citet{C73a}     &                     & \\
TYC 2394-539-1    & 11.61 & 0.22  &   0.12 & \citet{HJI61}    &                  & \\
TYC 2394-629-1    & 12.00 & 0.16  & -0.62 & \citet{MJD95}  & B1.5V            & \citet{MJD95}\\
TYC 2394-1141-1  & 11.04 & 0.44  & -0.46 & \citet{HJI61}    & B0.5V            & \citet{NMIB07}\\
TYC 2394-1214-1  & 10.15 & 0.51  & -0.49 & \citet{HJI61}    & O8               & \citet{HA65}\\
TYC 2394-1469-1  & 11.07 & 1.28  &  1.17 & \citet{C73a}      &                       & \\
TYC 2394-1500-1  & 10.70 & 0.30  & -0.62 & \citet{HJI61}    & B0.2V            & \citet{MJD95}\\
TYC 2394-1502-1  & 11.68 & 0.19  & -0.55 & \citet{MJD95}  & B1.5III            & \citet{MJD95}\\
TYC 2394-1594-1  & 11.86 & 0.37  &  0.22 & \citet{C73a}       &                    & \\
TYC 2394-1623-1  & 10.86 & 0.26  &           & \citet{HFM00} &                      & \\
TYC 2394-1691-1  & 11.34 & 0.26  & -0.64 & \citet{MJD95} & B0.5V             &  \citet{MJD95}\\
TYC 2394-1738-1  & 11.19 & 0.25  & -0.67 & \citet{HJI61}   & B0.5V              &  \citet{MJD95}\\
TYC 2394-1744-1  & 11.78 & 0.23  & -0.45 & \citet{HJI61}    &                    & \\
TYC 2394-1912-1  & 11.23 & 0.28  & -0.62 & \citet{HJI61}    & B0.5V            & \citet{MJD95}\\
TYC 2394-1933-1  & 11.21 & 0.19  & -0.63 & \citet{MJD95} & B1V                & \citet{MJD95}\\
\hline
\label{tab1}
\end{tabular}
\end{minipage}
\end{table*}

As the evolution of protostars and planet formation are closely related to circumstellar disc 
evolution, the mass accretion rate provides a useful diagnostic of the evolution 
of the circumstellar disc. Since the emergence of the magnetospheric accretion model 
\citep{US85,BBB88,K91} for PMS stars, several observational characteristics associated 
with accretion activities, such as blue/ultraviolet (UV) continuum excess emission, H$\alpha$ emission line 
profile, line luminosities of Br$\gamma$, Pa$\beta$, He {\scriptsize \textsc{I}} 5876, [O {\scriptsize \textsc{I}}] 
6300, etc., have been investigated (\citealt{GHBC98,CG98,MHC98,CMB04,NTM04,
MLBHC05,FBW09,MCM11}, and references therein). A correlation between the mass accretion 
rate ($\dot{M}$) and stellar mass ($M_{\mathrm{stellar}}$), $\dot{M} \propto M^\mathrm{b}_{\mathrm{stellar}}$ 
was found. The power law index $\mathrm{b}$ is in the range of 1 to 3 (\citealt{MLBHC05,NTR06,
FKB13} and references therein). However, the physical basis of the observed correlation between 
the mass accretion rate and the mass of the central star is not clearly understood. Furthermore, the mass accretion rate of 
intermediate-mass stars ($> 2 M_{\sun}$) seems to be quite uncertain. A few studies 
that have focused on Herbig Ae/Be stars found that the relation between $\dot{M}$ and 
$M_{\mathrm{stellar}}$ does not differ from that of low-mass stars \citep{CMB04,GNTH06,DB11}. 
On the other hand, some studies found a far steeper relation for Herbig Ae/Be stars with 
masses larger than $2 \ M_{\sun}$ \citep{MCM11,LSKBK14}. The accretion properties of these 
intermediate-mass PMS stars are one of the interesting issues.

The Sejong Open cluster Survey (SOS) project is dedicated to provide homogeneous 
photometric data down to $V \sim 20$ mag for many open clusters in the Galaxy. The overview of 
the project can be found in \citet[hereafter Paper 0]{SLB13}. Young open clusters NGC 2353 and 
IC 1848 were studied as a part of this project \citep{LSKI11,LSKBK14}. This paper on NGC 1893 is the fourth 
in the series. The observations and comparisons with previous photometry are described 
in Section 2. In Section 3, we discuss the reddening law in the direction of NGC 1893 and 
present fundamental parameters estimated from the photometric diagrams. The IMF of 
NGC 1893 and the mass accretion rates of PMS stars with UV excesses are presented 
in Section 4 and 5, respectively. Plausible triggering mechanisms within NGC 1893 are 
discussed in Section 6. Finally, we summarize the results from this study in Section 7.

\section{OBSERVATION}
The observation of NGC 1893 was made on 2009 January 19 using the AZT-22 1.5m 
telescope (f/7.74) at Maidanak Astronomical Observatory in Uzbekistan. All imaging 
data were acquired using the Fairchild 486 CCD (SNUCam; \citealt{IKC10}) 
with the standard Bessell $UBVI$ \citep{B90} 

\begin{landscape}
\begin{table} {\tiny
\caption{Photometric Data}
\label{tab2}
  \begin{tabular}{rcccccccccccccccccc}
  \hline
  \hline
ID$^1$ & $\alpha_{J2000}$  & $\delta_{J2000}$ & $V$ & $I$ & $V-I$ & $B-V$ & $U-B$ & $H-C$$^2$ &
$\epsilon_V$ & $\epsilon_I$ & $\epsilon_{V-I}$ & $\epsilon_{B-V}$ & $\epsilon_{U-B}$ & $\epsilon_{H-C}$ & N$_{obs}$ & 2MASSID &
H $\alpha$$^3$ & Sp$^4$\\
  \hline
 3421 & 05 22 53.62 & +33 22 52.5 & 21.383  & 18.863  &  2.520  &  1.895  &         &         & 0.040   & 0.004   & 0.040   & 0.131   &         &         & 1 2 1 1 0 0 &        -         & - &    \\
 3422 & 05 22 53.63 & +33 28 40.9 & 20.690  & 18.330  &  2.357  &  1.781  &         & -0.113  & 0.001   & 0.007   & 0.007   & 0.049   &         & 0.056   & 2 2 2 1 0 1 & 05225363+3328412 & - &    \\
 3423 & 05 22 53.64 & +33 23 53.3 & 20.279  & 17.551  &  2.728  &  1.902  &         &  0.045  & 0.001   & 0.017   & 0.017   & 0.038   &         & 0.064   & 2 2 2 1 0 1 & 05225364+3323532 & - &    \\
 3424 & 05 22 53.64 & +33 24 30.5 & 21.176  & 18.699  &  2.475  &  1.848  &         & -0.100  & 0.032   & 0.004   & 0.032   & 0.070   &         & 0.113   & 1 2 1 1 0 1 &        -         & - &    \\
 3425 & 05 22 53.64 & +33 25 49.2 & 21.557  & 18.901  &  2.654  &  1.645  &         & -0.035  & 0.043   & 0.041   & 0.060   & 0.104   &         & 0.122   & 1 2 1 1 0 1 &        -         & - &    \\
 3426 & 05 22 53.66 & +33 21 32.3 & 20.573  & 19.053  &  1.510  &  1.213  &         &         & 0.005   & 0.020   & 0.021   & 0.034   &         &         & 2 2 2 1 0 0 &        -         & - &    \\
 3427 & 05 22 53.66 & +33 28 16.2 & 20.770  & 19.058  &  1.698  &  1.333  &         &  0.115  & 0.021   & 0.017   & 0.027   & 0.051   &         & 0.097   & 1 2 1 1 0 1 &        -         & - &    \\
 3428 & 05 22 53.67 & +33 33 26.0 & 22.621  & 19.474  &  3.148  &         &         &         & 0.121   & 0.031   & 0.125   &         &         &         & 1 2 1 0 0 0 &        -         & - &    \\
 3429 & 05 22 53.69 & +33 29 08.6 & 16.100  & 15.407  &  0.689  &  0.524  &  0.342  &  0.421  & 0.004   & 0.001   & 0.004   & 0.006   & 0.006   & 0.007   & 2 2 2 2 2 2 & 05225368+3329086 & - &    \\
 3430 & 05 22 53.72 & +33 23 31.5 & 12.004  & 11.737  &  0.267  &  0.160  & -0.620  &  0.224  & 0.011   & 0.016   & 0.019   & 0.010   & 0.010   & 0.020   & 1 1 1 1 1 1 & 05225371+3323314 & - & B1.5V \\
 3431 & 05 22 53.73 & +33 28 25.8 & 22.584  & 20.316  &  2.269  &         &         &         & 0.088   & 0.053   & 0.103   &         &         &         & 1 1 1 0 0 0 &        -         & - &    \\
 3432 & 05 22 53.73 & +33 33 54.5 & 21.629  & 19.897  &  1.716  &  1.242  &         &         & 0.053   & 0.046   & 0.070   & 0.070   &         &         & 1 1 1 1 0 0 &        -         & - &    \\
 3433 & 05 22 53.74 & +33 25 09.9 & 22.334  & 20.861  &  1.473  &         &         &         & 0.065   & 0.087   & 0.109   &         &         &         & 1 1 1 0 0 0 &        -         & - &    \\
 3434 & 05 22 53.74 & +33 30 36.0 & 22.062  & 18.827  &  3.234  &  1.599  &         &         & 0.079   & 0.030   & 0.085   & 0.184   &         &         & 1 2 1 1 0 0 &        -         & - &    \\
 3435 & 05 22 53.76 & +33 27 59.3 & 20.430  & 18.977  &  1.442  &  1.162  &         &  0.043  & 0.026   & 0.003   & 0.026   & 0.040   &         & 0.131   & 1 2 1 1 0 1 &        -         & - &    \\
 3436 & 05 22 53.76 & +33 31 50.7 & 14.000  & 12.989  &  1.003  &  0.726  &  0.438  &  0.331  & 0.011   & 0.015   & 0.019   & 0.016   & 0.012   & 0.022   & 1 1 1 1 1 1 & 05225376+3331506 & - &    \\
 3437 & 05 22 53.79 & +33 31 27.4 & 22.441  & 20.303  &  2.138  &         &         &         & 0.088   & 0.073   & 0.114   &         &         &         & 1 1 1 0 0 0 &        -         & - &    \\
 3438 & 05 22 53.80 & +33 25 51.5 & 21.361  & 18.762  &  2.592  &  1.611  &         & -1.451  & 0.043   & 0.021   & 0.048   & 0.093   &         & 0.059   & 1 2 1 1 0 1 &        -         & H &    \\
 3439 & 05 22 53.80 & +33 27 56.5 & 22.687  & 20.531  &  2.157  &         &         &         & 0.086   & 0.064   & 0.107   &         &         &         & 1 1 1 0 0 0 &        -         & - &    \\
\hline
\end{tabular}
\begin{tabular}{@{}l@{}}
$^1$ The negative numbered ID represents the data from \citet{JM55,H56,HJI61,C73a,MJD95,HFM00} \\
$^2$ $H-C$ represents the H$\alpha$ index [$\equiv$ H$\alpha - (V+I)/2$] \\
$^3$ H: H$\alpha$ emission stars; h: H$\alpha$ emission star candidates \\
$^4$ Spectral type -- \citet{HA65,MJD95,NMIB07,SAW11} \\
\end{tabular}}
\end{table}
\end{landscape}

\noindent and H$\alpha$ filters. \citet{LSKI08} have 
described the characteristics of the CCD in detail. The mean seeing was better than 1.0 arcsec, and  
sky conditions were good. The observations comprised a total of 10 frames that
were taken in two sets of exposure times for each band -- 5s and 120s in $I$,  10s and 300s in $V$, 
20s and 600s in $B$, 30s and 600s in $U$, and 60s and 600s in H$\alpha$. We present the finder chart for the stars brighter 
than $V = 15$ mag in Fig.~\ref{fig1} using the Guide Star Catalogue version 2.3 \citep{LLM08}. 
The photometry for 24 stars that were saturated in the images was taken from previous studies. The 
photometric data and the adopted spectral types of the stars are presented in Table ~\ref{tab1}. 

\begin{figure*}
\includegraphics[height=0.45\textwidth]{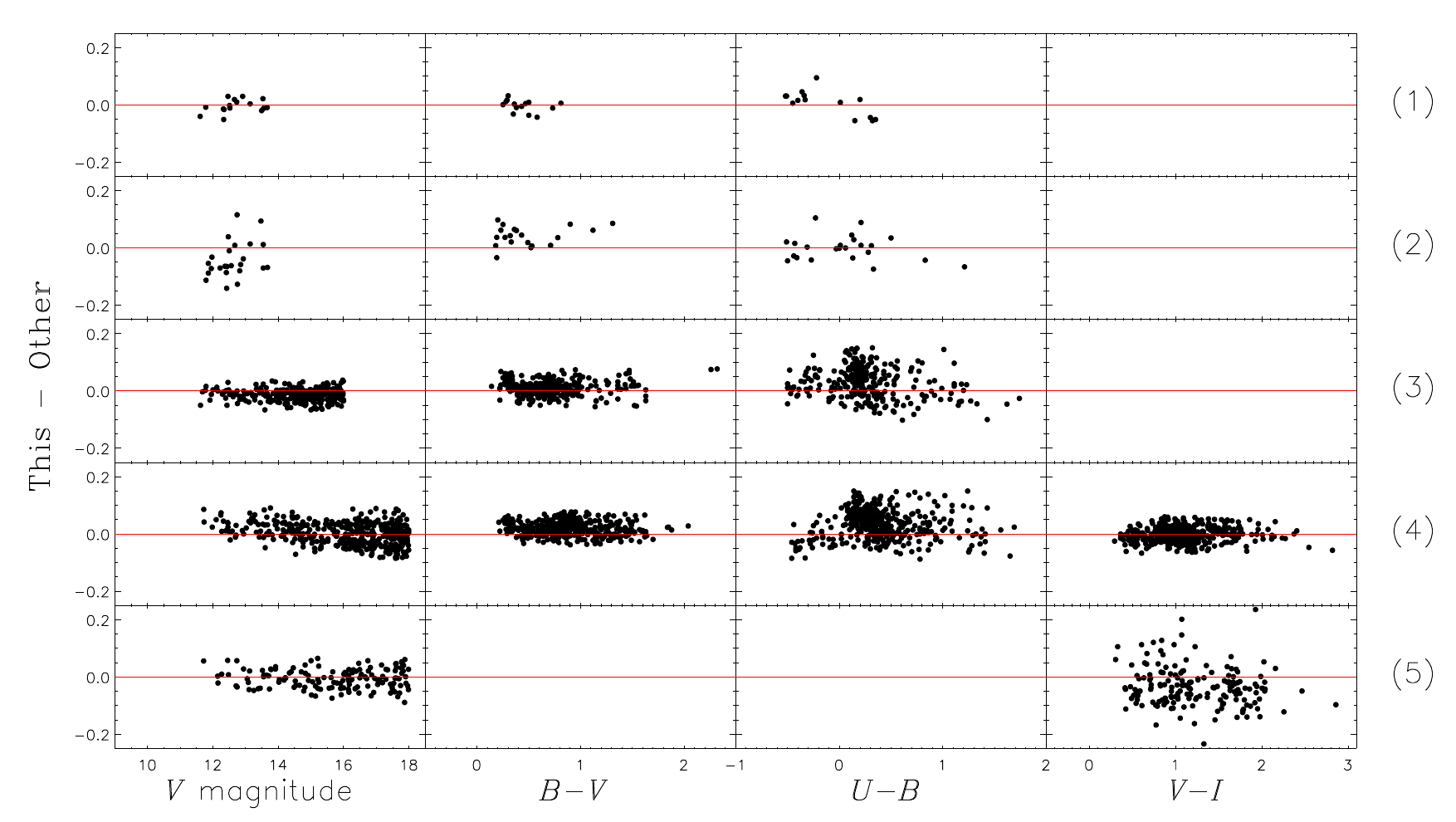}
\caption{Comparisons of our photometry with five previous sets of photoelectric and 
CCD photometry. Each panel from upper to lower represents the difference between us 
and (1) \citet{HJI61}, (2) \citet{C73a}, (3) \citet{MJD95}, (4) \citet{SPO07}, and (5) \citet{PSM11}, 
respectively. }
\label{fig2}
\end{figure*}

All pre-processing to remove instrumental artifacts was performed using the IRAF 
\footnote{Image Reduction and Analysis Facility is developed and distributed by the National 
Optical Astronomy Observatories, which is operated by the Association of Universities for 
Research in Astronomy under operative agreement with the National Science Foundation.}/CCRED 
packages as described in \citet{LSKI08}. In order to transform instrumental magnitudes to 
the standard magnitude and colours, we observed many equatorial standard stars 
\citep{MMLCE91} at different air masses. Since NGC 1893 was observed on the same night as 
the young open cluster IC 1848 \citep{LSKBK14}, the same daily parameters as used in the 
previous work (see table 2 in the paper), such as atmospheric extinction coefficients, photometric zero points, 
and time variation of the photometric zero points, were adopted in this work. We carried out the 
point spread function photometry for the target images using IRAF/DAOPHOT, and then the instrumental 
magnitudes were transformed to the standard magnitude and colours using the recently 
modified transformation relations for the SNUCam by \citet{LSBKI09} with the coefficients 
derived from the aperture photometry of standard stars. A detailed procedure of 
our photometry is delineated in \citet{LSKBK14}. Assuming that the luminosity function for 
the observed stars has a linear slope across the entire magnitude range, our photometry is 90 per cent complete 
down to $V = 20$ mag, which corresponds to $1 M_{\sun}$. It is worth noting that the completeness of 
our photometric data may be an upper limit because an obscuring cloud covers part of the cluster as seen in the 
right-hand side image of Fig.~\ref{fig1}. The photometric data for 6,503 stars from this 
work are available in the electronic table (Table ~\ref{tab2}) or from the authors (BL or HS).

We inspected the consistency of our photometry against previous studies. The photometric 
data in previous studies for NGC 1893 were obtained from the open cluster data base 
WEBDA\footnote{http://www.univie.ac.at/webda/}. The photometric data in this work were compared 
with those from 5 different studies (see Fig.~\ref{fig2}). In order to avoid the large photometric 
errors of faint stars, bright and isolated stars were used in the comparison ($V < 14$ mag for \citealt{HJI61}; 
\citealt{C73a}, $V < 16$ mag for \citealt{MJD95}; $V < 18$ mag for \citealt{SPO07,PSM11}). 
We present the mean and standard deviation of the differences between our photometry 
and that of previous studies in Table ~\ref{tab3}. Our photometric data are in good agreement 
with those of \citet{HJI61} (see also \citealt{LSKBK14}). The differences between the photometry 
of \citet{C73a} and this work appear to be slightly larger in $V$ and $B-V$, 
while the $U-B$ zero points are well consistent. Our photometry is in
good agreement with that of \citet{MJD95} and \citet{SPO07}. The $V-I$ colour of \citet{PSM11} 
reveals a large scatter, while a small scatter can be found in the comparison of our $V-I$ with that of \citet{SPO07}. 
Consequently, the photometric data in this work are well consistent with those of previous studies 
and is well tied to the Johnson-Cousins standard system. 

\begin{table*}
\begin{minipage}{180mm}
\caption{Comparison of Photometry}
\begin{tabular}{lcrcrcrcr}
\hline
\hline
Reference & $\Delta V$ & $N(m)$ & $\Delta (B-V)$ & $N(m)$ & $\Delta (U-B)$ & $N(m)$ & $\Delta (V-I)$ & $N(m)$ \\
\hline
\citet{HJI61}   &$-0.004\pm0.023$&16(2)    &$-0.004\pm0.021$&14(0)    &$0.007\pm0.044$  &14(0)&  &        \\
\citet{C73a}     &$-0.042\pm0.063$&24(6)    &$ 0.041\pm0.034$ &20(4)    &$-0.001\pm0.044$&23(1)&  &       \\
\citet{MJD95} &$-0.014\pm0.021$&263(43)&$ 0.010\pm0.027$ &277(22)&$ 0.028\pm0.053$ &280(18)&&\\
\citet{SPO07} &$ 0.002\pm0.037$ &414(42)&$ 0.019\pm0.024$ &423(30)&$0.035\pm0.050$  &405(44)&$-0.002\pm0.026$&408(42)\\
\citet{PSM11} &$-0.006\pm0.034$&140(31)&                              &             &                              &            &$-0.022\pm0.135$&169(2)  \\
\hline
\label{tab3}
\end{tabular}
\end{minipage}
\end{table*}

\section{PHOTOMETRIC DIAGRAMS}

Open clusters provide invaluable information on the stars in clusters, such as distance and 
age, which are difficult to obtain from field stars. These fundamental parameters are often 
determined in optical photometric diagrams using well-calibrated empirical 
relations and stellar evolutionary models. Because open clusters are mainly distributed in the 
Galactic plane, membership selection is essential to study these objects. In this section, 
we present the membership selection criteria, the reddening law, and fundamental parameters 
of NGC 1893 from two-colour diagrams (TCDs) in Fig.~\ref{fig3} and CMDs in Fig.~\ref{fig4}. 

\subsection{Membership Selection}
\begin{figure*}
\includegraphics[height=0.8\textwidth]{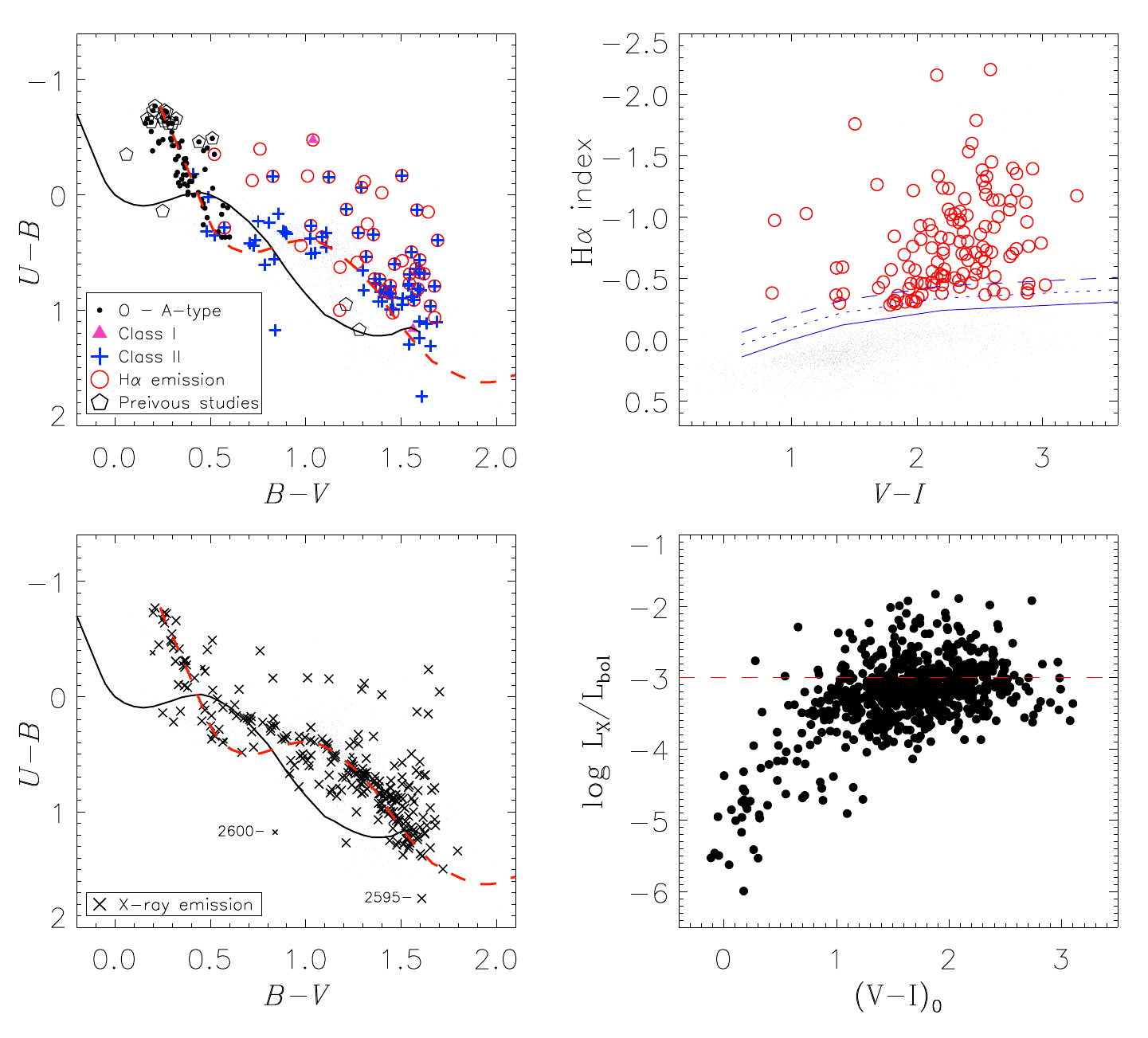}
\caption{Colour-colour diagrams of NGC 1893 and $\log L_{\mathrm{X}}/L_{\mathrm{bol}}$ of PMS members 
with respect to $(V-I)_0$. In the upper left-hand side panel, bold dots (black), triangles (magenta), 
pluses (blue), open circles (red), and open pentagons (black) represent early-type members (O -- A-type), 
Class I, Class II, H$\alpha$ emission stars, and bright stars obtained from previous studies, respectively. 
X-ray sources (large cross) and candidates (small cross) are shown in the lower left-hand side panel. The 
intrinsic and reddened colour-colour relations are overplotted by a solid and dashed line. The mean reddening of 
$\langle E(B-V) \rangle=0.56$ mag is adopted for the latter. In the upper right-hand side panel, 
the solid line represents the empirical photospheric level of unreddened MS stars, while the dotted and dashed 
lines are the lower limit of H$\alpha$ emission candidates and H$\alpha$ emission stars, respectively. 
From these criteria, 104 H$\alpha$ emission stars and 22 candidates are identified. Dashed line in the lower right-hand 
side panel denotes the saturation level of dynamo action.}
\label{fig3}
\end{figure*}

\begin{figure*}
\includegraphics[height=0.8\textwidth]{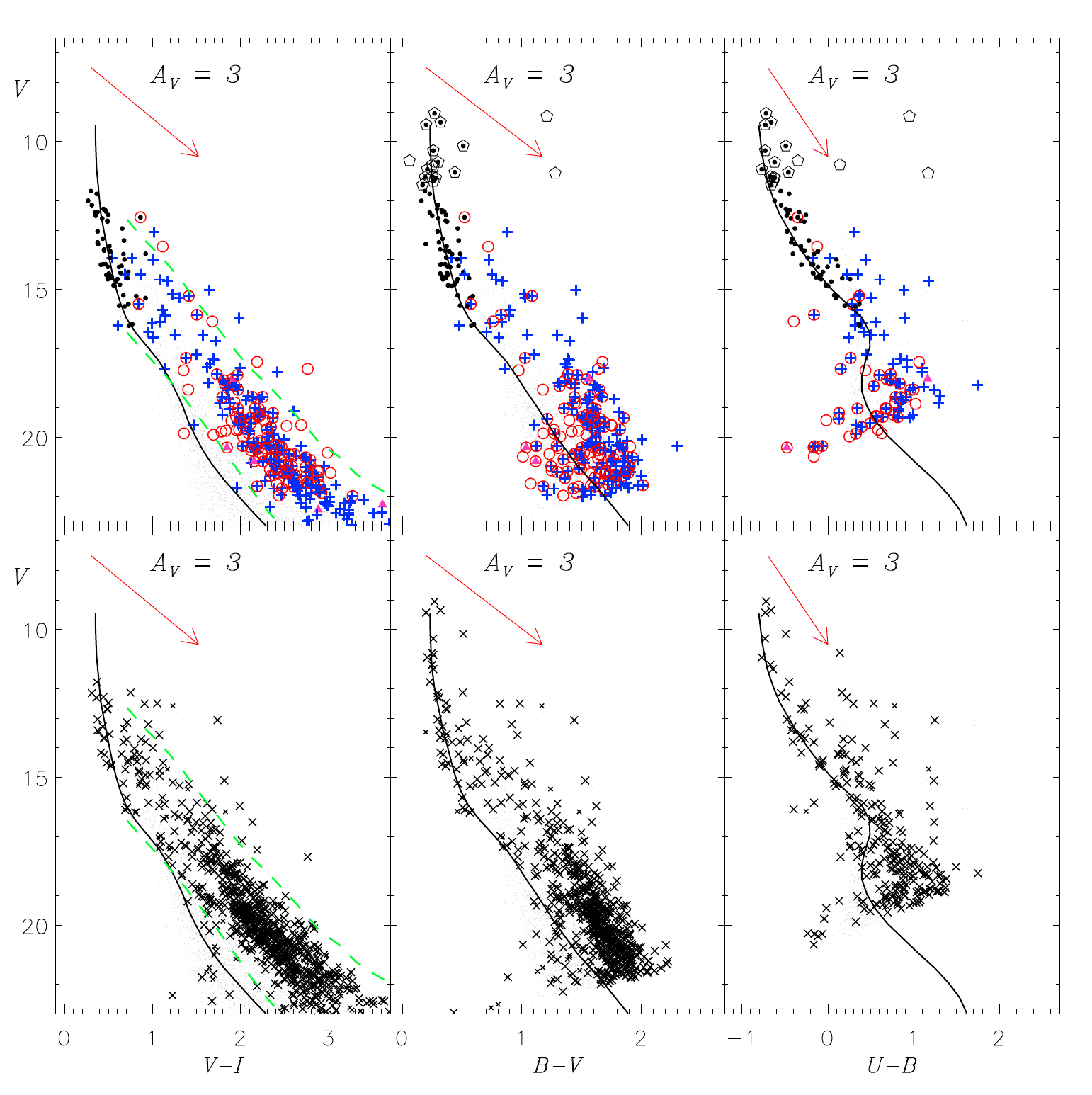}
\caption{Colour-magnitude diagram of NGC 1893. Left-hand side panels: $V-I$ vs. $V$ diagram. Dashed 
lines (green) denote the PMS locus. Middle panels: $B-V$ vs. $V$ diagram. Right-hand side panels: 
$U-B$ vs. $V$ diagram. The solid lines represent the reddened zero-age main sequence 
relation of \citet{SLB13}. The arrow denotes a reddening vector corresponding to $A_V = 3$ mag. 
The other symbols are the same as Fig.~\ref{fig3}. }
\label{fig4}
\end{figure*}

Early-type stars (O -- late-B) in either star-forming regions (hereafter SFRs) or young open clusters appear very prominently 
in photometric diagrams, especially in $U-B$, due to their high surface 
temperature. In addition, the individual reddening and distance moduli of the stars 
can be determined from the ($U-B, B-V$) TCD and CMDs, respectively, (see Section 3.2 and 3.3). 
The criteria for the early-type main sequence (MS) members are (1) $V \leq 16$ mag, $0.0$ mag 
$\leq B-V \leq 0.6$ mag, $-1.0$ mag $\leq U-B \leq 0.5$ mag (see left-hand side panel of 
Fig.~\ref{fig3} and Fig.~\ref{fig4}), $E(B-V) \geq 0.38$ mag, and $-1.0 \leq$ 
Johnson's $Q$ $\leq -0.2$, (2) an individual distance modulus between $\langle V_0 - M_V \rangle _{\mathrm{cl}}
- 0.75 - 2.5\sigma_{V_0 - M_V}$ and $\langle V_0 - M_V \rangle _{\mathrm{cl}} + \ 2.5\sigma_{V_0 - M_V}$ to 
take into account the effect of binary members (Johnson's $Q > -0.4$) and photometric errors 
\citep{SB99,KSB10,LSKI11,LSKBK14}, where $\langle V_0 - M_V \rangle _{\mathrm{cl}}$ and $\sigma_{V_0 - M_V}$ 
are the mean distance modulus and the width of the Gaussian fit to the distance modulus, 
respectively. Stars (ID 2576 and 4681) identified as Class II objects by \citet{CMP08} 
were excluded in this membership selection for early-type MS stars. These stars were 
assigned as PMS members. A total of 65 early-type MS stars were classified as members of NGC 1893.

X-ray, UV, H$\alpha$, and infrared (IR) excess emissions are known 
as good membership criteria for PMS stars. H$\alpha$ photometry has proven to be an 
efficient way to identify the PMS members in young open clusters ($\leq$ 3 Myr). Since 
\citet{SBL97} succeeded in selecting many PMS members of NGC 2264 using 
H$\alpha$ photometry, this technique has been widely used to distinguish PMS members 
from field interlopers along the line-of-sight to several young open clusters, e. g. NGC 6231 
\citep{SBL98,SSB13}, NGC 6530 \citep{SCB00}, NGC 2244 \citep{PS02}, NGC 2264 \citep{PSBK00,SBC04,SBCKI08}, 
NGC 3603 \citep{SB04}, and IC 1848 \citep{LSKBK14}. 
We found 105 H$\alpha$ emission stars and 22 candidates from the upper right-hand side panel of 
Fig.~\ref{fig3}. The star ID 826 ($V = 18.387$, $V-I = 1.404$, $B-V = 1.177$, and $U-B = 1.001$) 
with H$\alpha$ emission was excluded in the membership selection, because its observed 
$B-V$ and $U-B$ colours are similar to 
those of a foreground star. A total of 126 H$\alpha$ emission stars and candidates 
were classified as members of NGC 1893, one of which is an early-type member (ID 4957).

Dust continuum emission from the circumstellar discs of PMS stars can be detected 
at NIR and MIR wavelengths. Prior to 
\citet{CMP08} many studies for NGC 1893 attempted to search for PMS members using 
the NIR TCD \citep{VRCG99,MSBPB07,NMIB07,SPO07}. However, dust emission is, in general, 
more prominent in the MIR than in the NIR. \citet{CMP08} carried out extensive observations 
with the {\it Spitzer} space telescope and provided a reliable young stellar object (YSO) catalogue for Class I and 
Class II objects in NGC 1893. In addition, \citet{PSM11} found 1061 Class II objects 
based on their Q index with the photometric data from the optical to the MIR.  
792 of these objects were newly identified. However, 170 PMS stars have lower luminosities than 
the majority of the PMS population in the ($V, V-I$) CMD (see figure 4 of their paper). They 
interpreted the lower luminosity stars as PMS stars with edge-on discs and accretion activity. 
Given their photometric errors (Fig. ~\ref{fig2}), as well as the considerable 
overlap between the YSOs and field star population, most of them are more likely to be field stars, 
because it is difficult to distinguish YSOs from similarly reddened field stars using IR colours. 
Hence, only YSOs (Class I and Class II in \citealt{CMP08}) 
with a significant excess in the MIR are considered to be PMS members and will be used 
in the further analyses. In order to find the optical counterparts of the YSOs in the 
catalogue of \citet{CMP08}, we searched for stars within a matching radius of 
1.0 arcsec. Out of 213 identified optical counterparts, 5 YSOs are Class I, and the others are Class II objects. 

Had we only used H$\alpha$ photometry and the YSOs from MIR observations, the majority of WTTSs 
would be missed (e. g. \citealt{PSBK00,SCB00,PS02,LSKBK14}). Since the photometric properties 
of WTTSs are very similar to those of MS stars \citep{FMS99}, it is difficult 
to separate the members from field interlopers in TCDs or CMDs. The most efficient way 
to search for those PMS members without H$\alpha$ emission or MIR excess is to find 
X-ray emission stars, because PMS stars are well-known X-ray emitting objects. 
By using X-ray source lists from the {\it Chandra} or XMM-Newton observations 
almost complete lists of PMS members in NGC 2264 and NGC 6231 have been made
\citep{SRNGV06,SBCKI08,SSB13}. For NGC 1893, \citet{CMP08} made very deep X-ray observations 
($\tau _{\mathrm{exp}} \sim 440$ ks) with the {\it Chandra} X-ray Observatory and provided 
an X-ray source list (see table 4 in the paper). However, the X-ray source list is likely incomplete 
because they only presented the X-ray sources with counterparts in their MIR data. 
Their later follow-up study \citep{PSM11} independently used a full version of the X-ray 
source list from the observation \citep{CMP08} to select PMS members, identifying 415 
WTTSs. We also used the published X-ray source list \citep{CMP12} for
our membership selection. The optical counterparts of X-ray emission sources and 
candidates were searched for  with a matching radius of 1.0 arcsec. If a source with 
a redder $V-I$ colour, putatively a PMS star, was found within 1.5 arcsec, 
that was assigned as an X-ray emission source candidate. We found optical counterparts for 
724 X-ray emission sources and 59 candidates. 

CMDs in the lower panels of Fig.~\ref{fig4} [see also the (U-B, B-V) TCD in the lower 
left-hand side panel of Fig.~\ref{fig3}] show a large number of X-ray sources (crosses). In the 
$(V, V-I)$ CMD most of them are concentrated within the PMS locus (dashed lines), 
which have been modified from that used in a study for the young open cluster NGC 2264 
\citep{SBCKI08}. The PMS locus is a quite useful boundary with which to isolate PMS members 
in the CMD (see \citealt{SBCKI08,HSB12,SB10,SSB13,LSKBK14}) because many X-ray sources 
with neither H$\alpha$ emission nor IR excess emission that lie outside the PMS locus 
may be non-members or non-stellar objects (see also \citealt{SBCKI08,SSB09}). We made 
a visual inspection of the optical images for both the X-ray sources brighter and fainter 
than the PMS locus. The majority of the sources brighter than the PMS locus are likely foreground F -- M-type 
stars within 500 pc from the Sun, given the large sample of X-ray emitting objects identified 
from the {\it ROSAT} All-Sky Survey in \citet{AAC09}. Such objects,  with the exception of
 H$\alpha$ emission stars and/or Class II objects, were excluded from our membership selection. 

Most of the X-ray sources fainter than the PMS locus may not be bona-fide PMS stars with an edge-on 
disc, because dust in the disc would absorb X-ray photons with $E \geq 0.8$ keV \citep{D03}. O and 
early B-type stars are well-known X-ray sources \citep{LoW80,LuW80}. Indeed, 32 out of the 65 
early-type cluster members are X-ray emitters (see Fig.~\ref{fig4} and 7). Nevertheless, it is 
difficult to imagine that a large number of high-mass stars have been formed at high Galactic 
latitude, and therefore we do not anticipate OB-type background interlopers 
below the lower boundary of the PMS locus. It is more likely that X-ray emitting objects below the PMS locus 
are extragalactic sources as suggested by \citet{CMP08,CMP12}. 
Extended sources as well as ambiguous objects with a low signal-to-noise ratio were assigned 
as non-members from visual inspection. A few objects with either H$\alpha$ 
emission or MIR excess emission are included in the member list. 

\begin{figure*}
\includegraphics[height=0.35\textwidth]{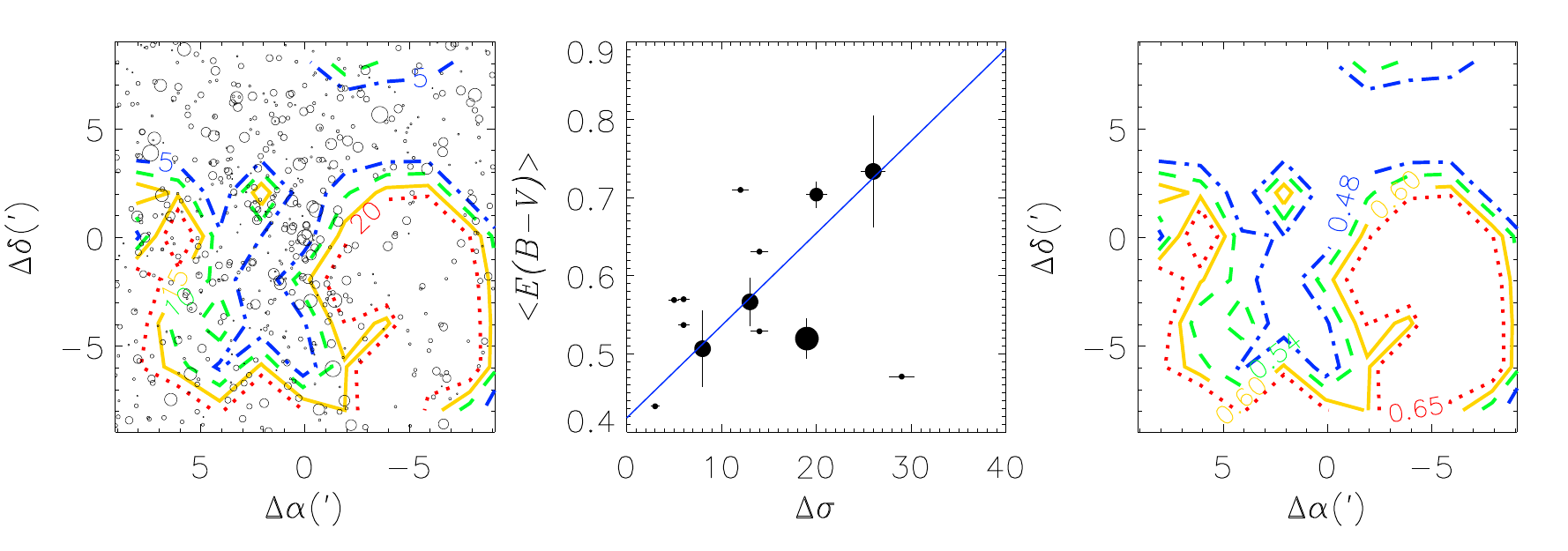}
\caption{Reddening map of NGC 1893. Left-hand side panel: spatial distribution of the observed stars 
in the $V$ band and contour map of the differential surface density $\Delta \sigma$, 
where $\Delta \sigma$ ($= \sigma_{[3.6]} - \sigma_V$) is the difference in the surface 
density of stars within a $2 \times 2$ arcmin$^2$ area in the 3.6 $\mu$m and $V$ bands. The 
field of view is the same as the solid box shown in the left-hand side panel of Fig.~\ref{fig1}. 
Dash-dotted (blue), dashed (green), solid (orange), and dotted (red) lines represent $\Delta \sigma 
=$ 5, 10, 15, and 20 per area, respectively. Middle panel: correlation between $\Delta \sigma$ and mean reddening 
$\langle E(B-V) \rangle$, where the mean reddening was obtained from the early-type members within a given 
$2 \times 2$ arcmin$^2$ area. The size of circles is proportional to the number of early-type members 
within a given area. The vertical bar represents the standard deviation of the reddening, while the 
Poisson statistics is assumed to be the error in the $\Delta \sigma$. The solid line represents an 
adopted relation, $E(B-V) = 0.012(\pm 0.002) \Delta \sigma$ + 0.418. Right-hand side panel: spatial 
variation of reddening. $\Delta \sigma$ in the left-hand side panel was converted to $E(B-V)$ using the 
relation in the middle panel. Dash-dotted (blue), dashed (green), solid (orange), dotted (red) lines represent $E(B-V) =$ 
0.48, 0.54, 0.60, and 0.65 mag, respectively.}
\label{fig5}
\end{figure*}

We also checked the position of those X-ray sources in the ($U-B$, $B-V$) TCD. A 
few outliers with significantly different photometric properties from the cluster sequence 
in both the CMDs and TCD were discounted as members. 
There are two Class II objects (ID 2595 and 2600) with an abnormal $U-B$ colour for their $B-V$ colour.
These stars fell upon a gap between two amplifiers of the Fairchild 486 CCD chip we used, 
where the gap is about 2 pixels ($\sim$ 0.53 arcsec) in width. 
The stars in the $V$ and $I$ band images were little affected by the column, while their profile was 
significantly distorted in the $U$ band image. Thus, we replaced the photometric data of the 
star ID 2600 with the mean values obtained in previous studies \citep{MJD95,SPO07}. 
In the case of the relatively faint star ID 2595, the photometric data of \citet{SPO07} did not include 
$U-B$, and so we only used our $V$ magnitude and $V-I$ colour in 
this work. There is a possibility that other field interlopers, such as FGK-type stars within $\sim$ 2 kpc from the Sun, 
still overlap with the PMS members within the locus. Using the reddening-corrected X-ray luminosity 
\citep{CMP12} and bolometric luminosity of the PMS members obtained from the Hertzsprung-Russell 
diagram (HRD; see section 3.4), we investigated the $\log {L_{X} / L_{\mathrm{bol}}}$ of such 
stars with respect to $(V-I)_0$ as displayed in the lower right-hand side panel of Fig.~\ref{fig3}. 
Stars later than mid-K clustered at the saturation level of dynamo action ($\log L_{X}/L_{\mathrm{bol}} = -3$), while the ratio 
between the X-ray and bolometric luminosities of stars with earlier spectral type rapidly declined with colour. There 
were no distinguishable field interloper amongst the stars selected as PMS members. However, we 
can not exclude the possibility of few foreground contamination with unusually high X-ray activity 
level which could not be discerned from the $\log {L_{X} / L_{\mathrm{bol}}}$ and $(V-I)_0$ relation. Another 
possible source of X-ray emission is accreting cataclysmic variables \citep{HG83}. \citet{DBP13} have 
discussed the detection chance of such variable stars. According to their simple estimation of the frequency, 
0.003 X-ray emitting cataclysmic variables are expected for a total of 6503 stars. Thus, the 
contribution of these stars to the field contamination is negligible.

\begin{figure*}
\includegraphics[height=0.45\textwidth]{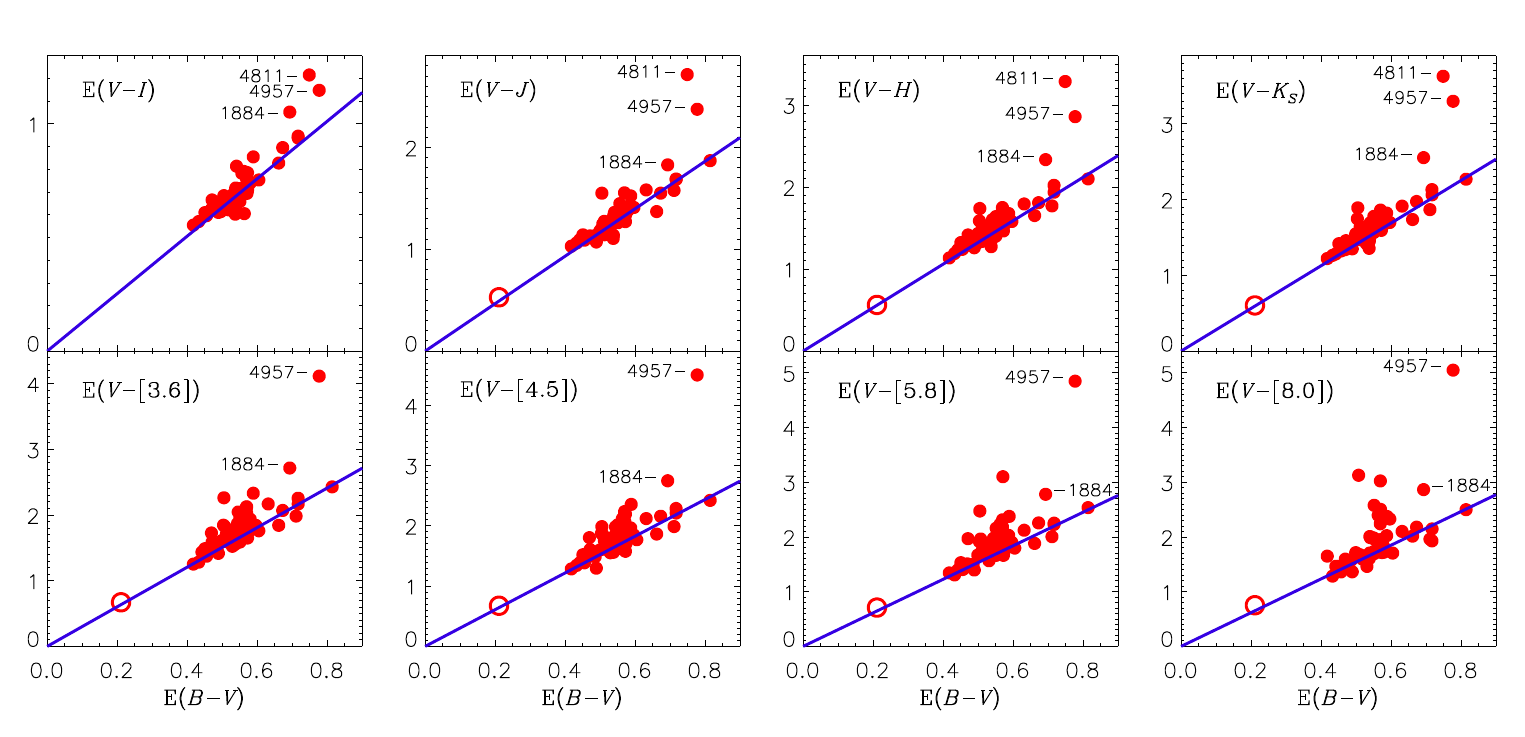}
\caption{Colour excess ratios obtained from the early-type members (bold dots) and 
a foreground early-type star (open circle). The solid line corresponds 
to $R_V = 3.1$. The colour excess ratios from the optical to the MIR data 
consistently show that the reddening law toward NGC 1893 is normal.}
\label{fig6}
\end{figure*}

In summary, 651 X-ray emission stars and 52 candidates were assigned to PMS membership
of NGC 1893. A total of 835 PMS members in NGC 1893 were selected using H$\alpha$ 
photometry, the YSO catalogue \citep{CMP08}, and the X-ray source list 
\citep{CMP12}. For the PMS members, we compared the detection efficiencies of the individual 
membership selection criteria within the completeness limit of 1 $M_{\sun}$. A 
total of 205 PMS members with mass larger than 1 $M_{\sun}$ were identified in this work. The 
PMS detection efficiencies from H$\alpha$ photometry, {\it Spitzer} MIR data, 
and {\it Chandra} X-ray data are about 10 per cent (21/205), 24 per cent (50/205), and 85 per cent (175/205), respectively. We 
found that only 9 PMS members exhibit X-ray, H$\alpha$, and MIR excess emissions 
simultaneously. The X-ray data appears to be the most efficient method of detecting young active 
stars, while H$\alpha$ and the MIR data show rather lower detection efficiency. 
These observational aspects may be related to the rapid evolution of 
circumstellar discs around higher mass PMS stars ($> 1 \ M_{\sun}$). The 
disc fraction of such stars within NGC 1893 is about 29 per cent. Given that 67 per cent of the PMS stars 
in this cluster have circumstellar discs \citep{CMP08}, H$\alpha$ and MIR data may become 
more efficient techniques for the selection of low-mass PMS stars \citep{SCB00,SSB09}. 

In addition, we considered the presence of A-type members which had already arrived at the ZAMS. 
These A-stars with significant X-ray emission are quite unusual, as apart from the peculiar (magnetic) A-type stars or the 
normal A-type star with a PMS companion \citep{PTPA99} most A-stars are not in X-ray catalogues. After adopting the mean reddening 
(see next section) for all stars, the criteria for such stars are (1) $V \leq 17$ mag, $0.0$ mag 
$\leq B-V \leq 0.6$ mag, -1.0 mag $\leq U-B \leq 0.6$ mag, $E(B-V) \geq 0.38$ mag, and 
Johnson's $Q > -0.2$, (2) an individual distance modulus between $\langle V_0 - M_V \rangle _{\mathrm{cl}} - 0.75 - 
2.5\sigma_{V_0 - M_V}$ and $\langle V_0 - M_V \rangle _{\mathrm{cl}} + \ 2.5\sigma_{V_0 - M_V}$, (3) $\mid \Delta 
(V_0 - M_V) \mid \leq 0.2 $ mag, where $\Delta$ means a difference between distance moduli 
derived from $B-V$ and $V-I$, and (4) no excess emission in the MIR and X-ray. The position 
of a few stars, which are compatible with those criteria, was checked in the ($U-B$, $B-V$) TCD. 
Only 6 probable A-type candidates were selected from these criteria. A total of 906 cluster 
members were used in the data analysis.
 
\begin{figure*}
\includegraphics[height=0.25\textwidth]{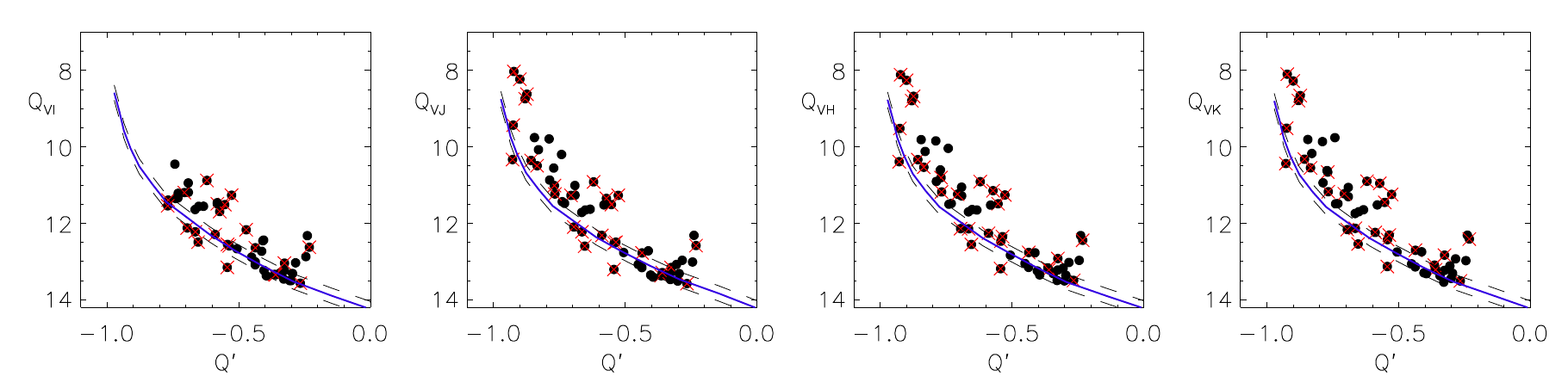}
\caption{Determination of the distance to NGC 1893. Crosses denotes X-ray emitting early-type 
members. The zero-age main sequence relations of \citet{SLB13} 
are used to determine the distance to the cluster after adjusting by $12.7 \pm 0.2$ mag, 
respectively. The solid line (blue) corresponds to the adopted distance of 3.5 kpc, and the 
dashed lines are ZAMS relations adjusted by the fitting errors.}
\label{fig7}
\end{figure*}

\subsection{Reddening and the Reddening Law}

The interstellar reddening toward young open clusters is, in general, determined by 
comparing the observed colours of early-type stars with the intrinsic ones 
in the ($U-B$, $B-V$) TCD along the reddening slope. Paper 0 provided reliable 
intrinsic colour relations (see table 1 in the paper). We used the relations to determine the 
individual reddening $E(B-V)$ of the early-type members of NGC 1893. Since the 
dependence of the reddening slope on $E(B-V)$ can be assumed negligible for less reddened 
stars [$E(B-V) < 1$ mag], we simply adopted $E(U-B)/E(B-V)=0.72$. 
For 63 isolated early-type members, a mean reddening was estimated to be 
$\langle E(B-V) \rangle=0.563 \pm 0.083$ (s.d.) mag and shown in the left-hand side panels of Fig.~\ref{fig3} 
as a dashed line. This value is well consistent with that of previous studies, e.g. 
$E(B-V) = 0.59$ mag \citep{JHIMH61}, 0.56 mag \citep{B63}, 0.59 -- 0.65 mag 
\citep{HA65}, 0.55 mag \citep{M72}, 0.53 mag \citep{MJD95}, 0.60 mag \citep{LGM01,PSM11}, 
and 0.4 -- 0.6 mag \citep{SPO07}. 

As mentioned in several previous studies \citep{TCER91,SPO07,PSM11}, a non-negligible amount of differential 
reddening across the observed field was found in the ($U-B$, $B-V$) TCD. Because 
this is very crucial for the reddening correction of PMS stars without spectral 
types, the spatial variation of the reddening was investigated using the position 
and reddening of early-type members as shown in various studies \citep{SPO07,HSB12,
SSB13,LCS13,LSKBK14}. However, the spatial distribution of early-type cluster members 
does not cover the western part of the observed region. In this paper, we 
introduce a new way to construct a reddening map. The basic principle is that 
MIR photons are less affected by the effect of the interstellar reddening than optical 
photons. The deep MIR photometric data of \citet{CMP08} 
provide a good reference against our optical data in the surface density of stars. The difference 
between the surface densities at MIR and optical wavelengths could indicate the amount of interstellar 
reddening. The larger the difference, the higher the reddening is in the line of sight. Since 
among the 4 IRAC bands, \citet{CMP08} detected the largest number of sources in the 3.6 $\mu$m band, 
we assigned the 3.6 $\mu$m data as the reference. In order to construct the differential 
surface density map we counted the number of stars observed in the $V$ and 3.6 $\mu$m 
bands within a given $2 \times 2$ arcmin$^2$ area. After subtracting the surface 
density in the $V$ band from that in the 3.6 $\mu$m band, we obtained the differential surface density 
map as shown in the left-hand side panel of Fig.~\ref{fig5}, which shows the same 
field of view as the observed region (solid box in the left-hand side panel of Fig.~\ref{fig1}). 
The figure shows that the differential surface density appears to be highest in 
the directions to the south and the south-west of the cluster centre. Indeed, the 
contour map is almost the same as the spatial distribution of the dark clouds 
in the right-hand side image of Fig.~\ref{fig1}. 

The differential surface density map can be converted into a reddening map by using 
the correlation between the surface density and the mean reddening of the early-type 
members within the same area, as shown in the middle panel of Fig.~\ref{fig5}. The 
error of the vertical axis represents the standard 
deviation of the reddening, while that of the horizontal axis was assumed to follow 
Poisson statistics. The smallest value [$E(B-V) = 0.418$ mag] among the reddening 
values of the isolated early-type members was considered as the intercept of the correlation. 
The slope was estimated using a $\chi ^2$ fitting method, where the standard 
deviation of the reddening within a given area was used as the uncertainty of a 
measurement. Thus, areas containing only one early-type member were excluded 
in the estimation of the slope. We obtained a correlation of $E(B-V) = 0.012(\pm 0.002)
\Delta \sigma \ + 0.418$, and that is compatible with the global trend of the data points. 
In the end, the reddening map of NGC 1893 was constructed 
from the differential surface density map through that relation. We present the resultant 
reddening map in the right-hand side panel of Fig.~\ref{fig5}. A typical uncertainty in 
$\langle E(B-V) \rangle$ estimated from the reddening map is about 0.1 mag if the error 
in the differential surface density is about 10. The reddening map 
was used to correct the individual reddening of the PMS members as well as the 
A-type MS members.

The reddening law is a useful tool to understand dust evolution in SFRs, 
because it is closely related to the size distribution of dust grains. For several 
extremely young SFRs, the ratio of total-to-selective extinction 
($R_V$) appears to be larger than that of the normal reddening law ($R_V = 3.1$) (see 
table 3 in \citealt{G10}). Previous photometric survey programs \citep{JHIMH61,HA65,
WH68,H78} have assumed the reddening law of most open clusters to follow the normal 
law. For NGC 1893, \citet{TCER91} studied the reddening law 
toward the cluster and found $R_V= 2.8$. This value is rather lower than 
the normal one. On the other hand, other studies \citep{NMIB07,SPO07}, which used 
colour excess ratios, confirmed that the reddening law of NGC 1893 is normal. A 
polarimetric study \citep{EPM11} also supported the normal reddening law toward 
the cluster. To check the previous results we investigated the reddening law toward 
NGC 1893 using various colour excess ratios as shown in the series of studies 
\citep{KSB10,LSKI11,HSB12,SSB13,LSKBK14}.

The NIR 2MASS data and MIR {\it Spitzer} IRAC data of \citet{CMP08} allow us to test 
the reddening law toward NGC 1893 in a consistent way from the optical to the 
MIR. The observed colours of 63 isolated early-type members 
were compared with the intrinsic colours for $V-I$ from table 1, 2MASS from 
table 2 in Paper 0, and {\it Spitzer} IRAC bands (Sung et al., in preparation) 
in order to compute the colour excess ratios $E(V-\lambda)/E(B-V)$ as shown in 
Paper 0, \citet{SSB13}, and \citet{LSKBK14}. We present the colour excess ratios 
in Fig.~\ref{fig6}. The solid line corresponds to the normal reddening law 
$(R_V = 3.1)$. Most members (bold dots) as well as a foreground early-type star (open circle) 
lie on the solid line, except for a few stars (ID 1884, 4811, and 4957). This result 
also supports the reddening law toward NGC 1893 (foreground as well as the 
intracluster medium of NGC 1893) being normal. The total extinction 
$A_V$ can be determined from $A_V = R_V \times E(B-V)$. From the individual reddening 
of early-type members obtained above, the mean total extinction of the stars was 
$\langle A_V \rangle = 1.74\pm 0.26$ mag.
 
\citet{NMIB07} found that a few late-B-type members of NGC 1893 deviated 
from the normal reddening law. We investigated the observed characteristics of 
the stars with excess emission in Fig.~\ref{fig6}. The star 
ID 1884 was identified as an X-ray emission star. Although the star was not 
identified as a Class I or a Class II object, the star shows excess emissions 
in all wavelengths. Another star ID 4811 with X-ray emission is 
located within the head of the emission nebula Sim 130. We guess that the star was 
likely to be missed in the IRAC images due to the bright nebula. The other star 
(ID 4957) without an X-ray emission was identified as an H$\alpha$ emission star, and 
it is located in the vicinity of Sim 130. All these facts indicate that the stars may be 
young high-mass stars. Besides these stars, there are a few stars with excess 
emission in the [5.8] and [8.0] bands. The excess emission may be due to the dust 
emission in the immediate of the star rather than to a variation of the reddening law 
in the intracluster medium. 

\subsection{Distance to NGC 1893}

The location and distance of young open clusters are very useful in tracing 
the local spiral arm structure of the Galaxy. The distance to open clusters can be determined 
by using the ZAMS fitting method. The reddening correction used to be crucial in obtaining a reliable 
distance. Paper 0 introduced a reliable ZAMS fitting method based on reddening-independent 
indices as below: 

\begin{equation}
 Q^{\prime} \equiv (U-B) - 0.72(B-V)-0.025E(B-V)^2 
\end{equation}
\begin{equation}
 Q_{VI} \equiv V - 2.45(V - I) 
\end{equation}
\begin{equation}
 Q_{VJ} \equiv V -1.33(V -J) 
\end{equation}
\begin{equation}
 Q_{VH} \equiv V - 1.17(V -H) 
\end{equation}
\begin{equation}
 Q_{VK_{S}} \equiv V - 1.10(V - K_{S})
\end{equation}

\noindent The indices are combinations of $V$ and $IJHK$ magnitudes respectively. 
Since the spectra of hot MS stars (O -- B-type) in 
the optical and NIR passbands do not exhibit strong lines 
apart from H, He, and a few light elements \citep{LR92,TW93,DBJ96,HCR96,P98} we believe that each index 
is insensitive to any metallicity effect. We determined the distance to NGC 1893 using 
the ZAMS fitting method with the reddening-independent indices. According to Paper 0, the 
ZAMS relation should be fitted to the lower ridge line of the MS band to avoid the effects of multiplicity and 
evolution as presented in Fig.~\ref{fig7}. We adjusted the ZAMS relation above and below 
the distribution of the early-type members in the $Q_{V\lambda}$-$Q^{\prime}$ planes and 
obtained a distance modulus of $V_0-M_V = 12.7 \pm 0.2$, equivalent to $3.5 \pm 0.3$ kpc. 
The uncertainty in the distance modulus was from Fig.~\ref{fig7}. This places NGC 1893 ($l = 173\fdg585$, 
$b = -1\fdg680$) in the Perseus spiral arm.
 
The distance moduli obtained from the $UBV$ photometry are broadly consistent with each other. 
Our result, in particular, is in good agreement with the results of previous studies 
($V_0 - M_V = 12.5$ mag, \citealt {H78}; 12.6 mag, \citealt{SPO07}, 12.8 mag, 
\citealt{B63,C73b,LGM01,PSM11}) within the uncertainty. Previous studies 
determined distances through the spectroscopic parallax, isochrone 
fitting, and the ZAMS fitting method. The spectroscopic parallax is an 
apparently easy way to determine the distance and reddening. However, 
it is difficult to take into account the evolution effects of early-type stars 
on the variation of their absolute magnitude. Furthermore, the intrinsic scatter in
the $M_V$ versus spectral-type diagram of massive evolved stars is large 
(see figure 3 in \citealt{HD79}). Therefore, the spectroscopic parallax 
can give distances slightly shifted from the true one in the process of averaging the 
distances of individual stars (e. g. \citealt{H78,MJD95}). In order to fit an 
isochrone to observed CMDs, at least 3 free parameters, such as reddening, 
distance, and age (metallicity if possible), are needed. While reddening can be easily 
determined if multi colour photometry including $U-B$ is available, it is difficult 
to constrain the age of young clusters ($<$ 10 Myr) unless the clusters host massive evolved 
stars at a specific evolutionary stage, such as yellow hypergiants, red supergiants, or Wolf-Rayet 
stars, as shown in several studies for the starburst cluster Westerlund 1 \citep{CNCG05,CHCNV06,
NCR10,LCS13}. Despite such complexities, the results obtained from isochrone fitting 
in previous studies \citep{LGM01,SPO07,PSM11} give a consistent value. A 
few studies \citep{JHIMH61,B63,M72} used the traditional ZAMS fitting method and 
obtained distance moduli of 12.8 and 13.0 mag, respectively. 
These results are roughly consistent with ours within the uncertainties. The difference 
between the ZAMS relations of \citet{J57} and Paper 0 is about -0.09 mag at $U-B = -0.4$ 
to -0.47 mag at $U-B = -0.9$. The slight discrepancy among the studies may result from 
the differences of the ZAMS relations used. 

On the other hand, the distances (4.4 -- 6.0 kpc) determined from Str\"{o}mgren $uvby$ 
photometry \citep{TCER91,F93,MBN01} appear to be systematically larger than those derived 
from $UBV$ photometry. The main targets of the studies were limited to bright stars in the NGC 1893 
field due to the narrow band width of the Str\"{o}mgren system. The error in the $b-y$ colour 
rapidly increased at $y =$ 13 -- 14 mag. These large errors likely influence the result of the 
ZAMS fitting. In addition, while the photometric data of \citet{TCER91} and \citet{F93} show a good 
consistency with each other, those of \citet{MBN01} reveal a systematic difference and a large scatter. 
The $b-y$ colour of \citet{MBN01} is $\sim$ 0.03 mag bluer than that of \citet{F93} at $b-y < 0.25$ mag. 
The maximum difference in the $m_1$ index is up to 0.1 mag, and the $c_1$ index shows a 
spread of 0.2 mag. Hence, the discrepancies and errors in the colours have most likely led to the 
overestimated distance (6.0 kpc). 

\begin{figure}
\includegraphics[height=0.42\textwidth]{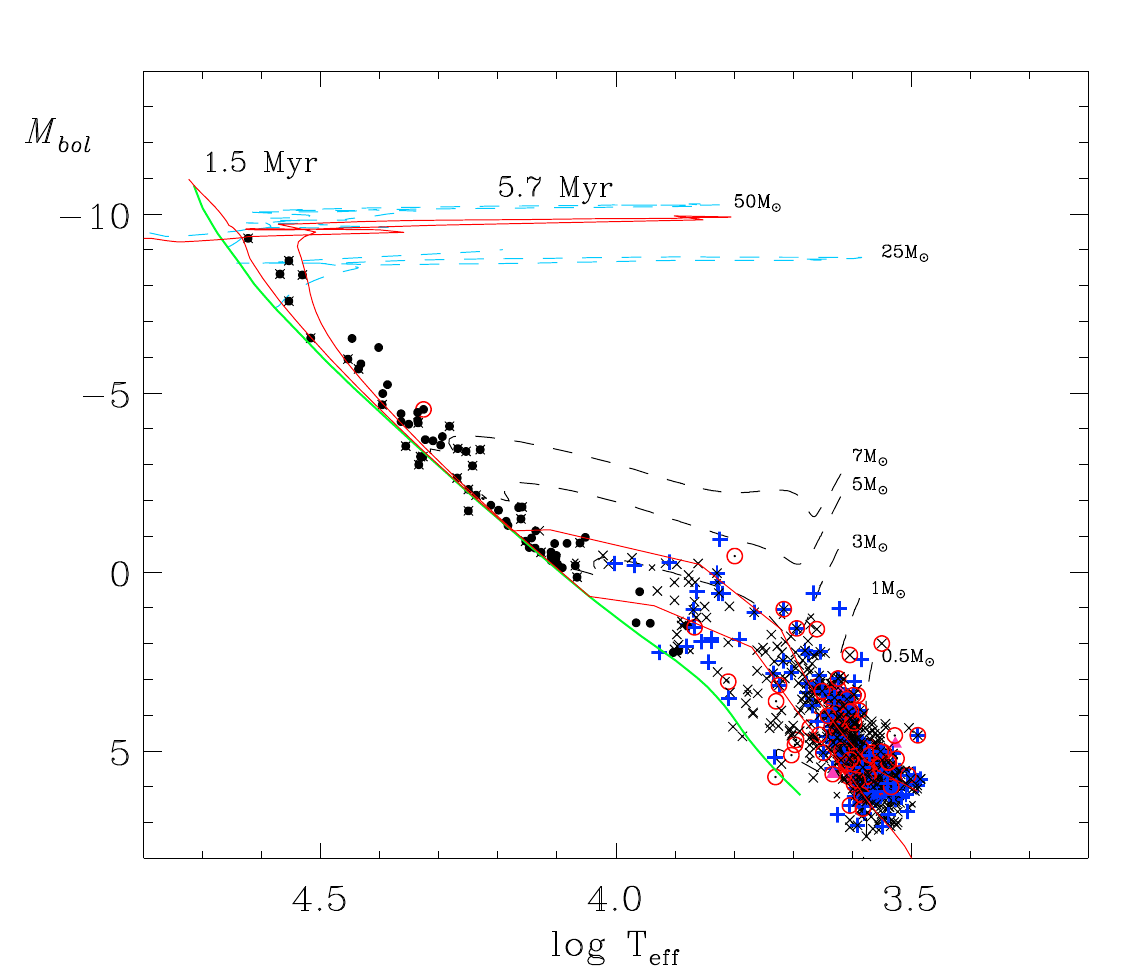}
\caption{The Hertzsprung-Russell diagram of NGC 1893. Isochrones (0., 1.5, and 5.7 
Myr) are superimposed on the diagram with several evolutionary tracks \citep{EGE12,SDF00}. 
The other symbols are the same as Fig.~\ref{fig3}. }
\label{fig8}
\end{figure}

\subsection{Hertzsprung-Russell Diagram and Age of NGC 1893}

We corrected for the reddening of the individual early-type members as described in 
the previous section. However, as it is very difficult to estimate reliable reddenings of PMS stars 
without accurate spectral types, the reddening map from Fig.~\ref{fig5}
was used in the reddening correction for the PMS members and A-type MS members. Applying the distance 
modulus of 12.7 mag to all members, the observed CMDs were transformed to the dereddened 
CMDs, $M_V$ against $(V-I)_0$, $(B-V)_0$, and $(U-B)_0$, and the HRD of 
NGC 1893 was then constructed using the several relations adopted in Paper 0.

The effective temperature of stars earlier than O9 was obtained from the spectral 
type-effective temperature relation from table 5 in Paper 0. For the temperature scale 
of the PMS members with $(V-I)_0 \leq 1.4$, the relation between $(V-I)_0$ and effective 
temperature of \citet{BCP98} was adopted, and another relation between $(V-I)_0$ and 
temperature \citep{B95} was applied to the remaining cool PMS members. For the 
other MS members, we averaged the temperature derived from spectral type-effective 
temperature relation and colour-temperature relations with an appropriate 
weight. The bolometric corrections for all the members were carried out using 
table 5 in Paper 0. We present the HRD of NGC 1893 in Fig.~\ref{fig8}. We note 
that 4 B-type stars were classified as giants by \citet{MJD95}. \citet{NMIB07} 
presented the more reliable spectral type and luminosity class for 2 B-type giant 
stars (HDE 243018 and TYC 2394-1141-1). We adopted their classifications in 
this work (see Table~\ref{tab1}). The luminosity class of two B-type stars, TYC 2394-450-1 
(B1.0III, $M_{\mathrm{bol}} = -4.988$, $\log T_{\mathrm{eff}} = 4.395$) and 
TYC 2394-1502-1 (B1.5III, $M_{\mathrm{bol}} = -4.460$, $\log T_{\mathrm{eff}} = 4.336$), 
is needs to be reexamined because \citet{MJD95} classified them as giants, but they 
are placed near the ZAMS line. If such stars are bona-fide giants, they may be an old 
population in the Auriga OB2 association. Although we adopted the published luminosity class (III) in 
this work it does not cause serious problems because the difference in temperature between 
a MS and giant star at B1.0 and B1.5 is about 500 -- 600 K and the difference 
in bolometric correction is only 0.01 -- 0.02 mag.  

Assuming solar composition, we roughly constrained the age of the cluster 
from the presence of an O4.5V star (HDE 242908). \citet{M13} summarized 
the properties of massive stars, such as the initial mass, effective temperature, 
luminosity, surface gravity, with spectral type. The lifetime of an O4V star is estimated 
to be 5.7 Myr. But with the presence of 3 O-type ZAMS stars (HDE 242935, 
BD +33 1025, and HDE 242926; \citealt{SAW11}), the age of the cluster may be younger 
than 5.7 Myr. Isochrone fitting to the HRD is a convenient way to estimate the 
age of a cluster. A few evolutionary tracks (dashed lines) with different initial masses 
and isochrones (solid lines) for different ages are superimposed on the HRD. 
The stellar evolutionary models of \citet{EGE12} taking into account the effect of 
stellar rotation on the evolution of stars ($Z = 0.014$) were used for MS stars, 
while we used the models of \citet{SDF00} for PMS stars. The isochrones (0.0, 
1.5, and 5.7 Myr) were constructed from the two evolutionary models. The lower boundary of 
the most massive stars in the HRD is well predicted by the isochrone with 
an age of 1.5 Myr. Thus, the MS turn-off age of NGC 1893 is about 1.5 Myr.

There are two previous studies for the light element abundance of NGC 1893 (e.g. 
\citealt{RBDF93,DC04}). \citet{RBDF93} have estimated the abundance of 6 stars using 
LTE models of stellar atmosphere and concluded that there is no evidence for low 
abundance. On the other hand, \citet{DC04} investigated the light element abundance 
of 2 cluster members with modern non-LTE models. The light elements were $\sim 0.26$ dex 
less abundant than those of clusters in the solar neighbourhood. If the result of 
the former is valid, the turn-off age in this work is likely appropriate. The Geneva 
group has published grids of stellar evolutionary models for extremely different 
chemical compositions ($Z = 0.014$, \citealt{EGE12}; $Z = 0.002$, \citealt{GEE13}). 
According to the models for high-mass stars in the range of 32 -- 60 $M_{\sun}$, the 
evolutionary tracks for a given mass are far different because of the large difference 
in chemical composition. We simply compared the age and luminosity 
of the high-mass stars at the MS turn-off in the solar metallicity models with 
those of their counterparts in the low-metallicity models. The solar metallicity 
models give an older age than low-metallicity models for the same mass. The differences 
between stars with different chemical composition were about 0.4 Myr 
in age and 0.1 mag in bolometric magnitude, respectively. We also considered the 
evolutionary models of \citet{BMC11}, in which they published grids of stellar 
evolutionary models for 3 different initial metallicity environments - the Galaxy, the 
Large and Small Magellanic Clouds. The isochrone with the chemical composition 
of the Galaxy for 1.8 Myr appears to well fit to the position of high-mass members 
in the HRD, while the models with the composition of the Large Magellanic Cloud 
gives an age of 2.3 Myr. Thus, a systematic uncertainty of age caused by the 
uncertainty of chemical composition is about 0.5 Myr in the case of very young 
open clusters ($<$ 2 Myr). 

\begin{figure}
\includegraphics[height=0.45\textwidth]{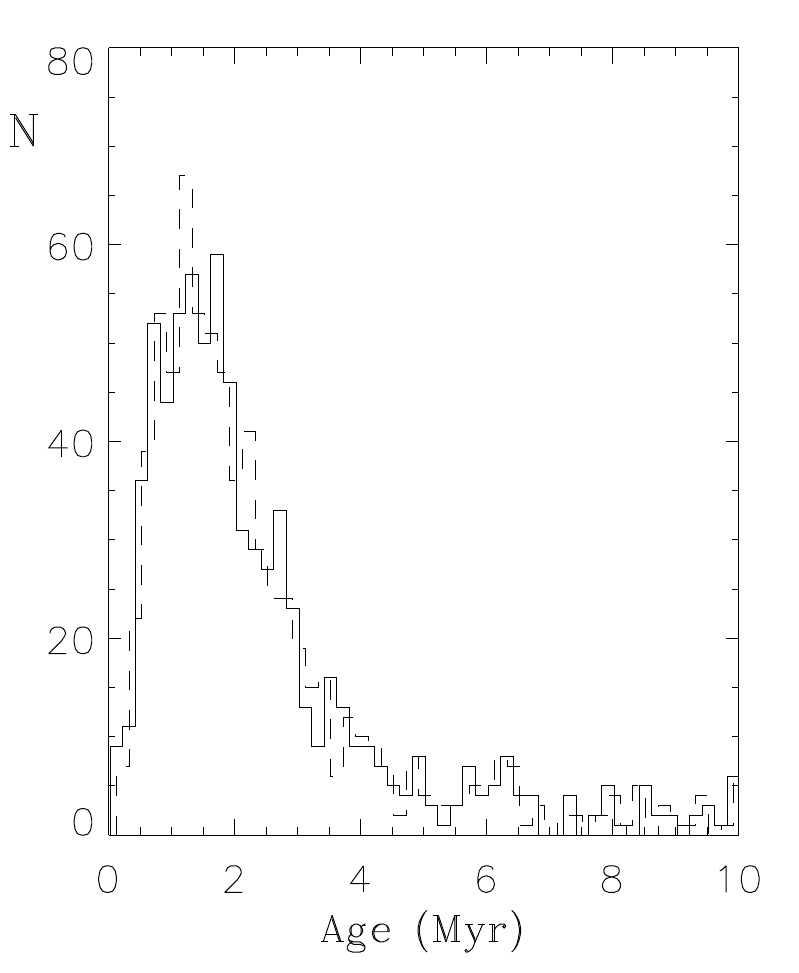}
\caption{Age distribution of PMS members. The median age is about 1.9 Myr with a 
spread of 5 Myr. The ages of the PMS members appear to be an asymmetric 
distribution with a long tail toward older age.}
\label{fig9}
\end{figure}

Although HDE 242926 is an O7Vz star according to \citet{SAW11}, 
its luminosity ($M_{\mathrm{bol}} = -8.70$) at a given effective temperature is higher 
than that of HDE 242935 ($M_{\mathrm{bol}} = -8.33$, O6.5V) and BD +33 1025 ($M_{\mathrm{bol}} = 
-7.57$, O7Vz). \citet{J72} has tagged the star as having a variable radial velocity, and therefore 
it is still possible that HDE242926 may be a binary/multiple system. The luminosity of the 
O8 star TYC 2394-1214-1 is also comparable to that of HDE 242935. We could not 
find the luminosity class of the star in \citet{HA65} as well as information on possible 
multiplicity, and therefore the star was assumed to be MS in this work. It is
possible that the star has more than one high-mass companion, or that its 
spectral type is incorrect. 

The majority of the PMS members have masses smaller than $3 M_{\sun}$ as seen in the 
HRD. Intermediate-mass PMS stars ($\sim 3 M_{\sun}$) seem to be approaching the 
ZAMS. There are quite a few PMS members that are near or on the ZAMS. These stars may be either PMS stars with an edge-on 
disc as discussed by \citet{PSM11} or old population stars in the Auriga OB 2 association. 
We also estimated the age of NGC 1893 from the PMS members using the PMS evolutionary 
models of \citet{SDF00} and present the age distribution in Fig.~\ref{fig9}. The peak of 
the distribution appears at 1.0 -- 1.5 Myr. The median age is about 1.9 Myr with an 
age spread of 5 Myr, where the spread was defined as the age difference between the 10 and 90 
percentiles in the cumulative age distribution of the PMS members \citep{SB10}. The age scale between 
the MS and PMS members also shows a good consistency. The age spread is very 
similar to that found in other young open clusters, such as NGC 2244 \citep{PS02}, 
NGC 2264 \citep{PSBK00}, NGC 6530 \citep{SCB00}, Trumpler 14 and 16 \citep{HSB12}, 
and IC 1848 \citep{LSKBK14}, being in a good agreement with that of \citet{PSC13} for NGC 1893.

The age spread of PMS stars is thought to be a key to understanding the formation processes 
of star clusters. However, there are well-known caveats for the derived age spread of 
PMS stars. A few observational uncertainties affect the luminosities and effective 
temperature of PMS stars. The inclusion of field interlopers in the member list can lead the age 
distribution to be wider \citep{H03}. We obtained the ages of PMS stars based on the carefully 
selected members (see Section 3.1). The influence of non-members on the age distribution may therefore be negligible. 
Differential reddening is, in general, found in extremely young open clusters and SFRs. 
Imperfect reddening correction for such PMS stars is one of the probable sources causing 
an apparent spread in luminosity and temperature. In addition, PMS stars occupy the faint 
part of CMDs, and so photometric errors are more likely to contribute to the spread of these stars.

We examined these issues using Monte-Carlo simulations. A total of 800 artificial 
PMS stars in the mass range of 0.5 to 5 $M_{\sun}$ were generated by using the PMS models 
of \citet{SDF00}. The underlying IMF was assumed to be the Salpeter IMF \citep{Sp55}. The age of the artificial 
stars was set to 2 Myr, and an instantaneous star formation history was assumed, i.e. no 
spread in the age. The luminosity and temperature were transformed to $V_0$ magnitude 
and $(V-I)_0$ color by using the relations of \citet{B95} and \citet{BCP98}. The reddening 
was assumed to be the mean value of $\langle E(B-V)\rangle = 0.56$ mag. In order to 
reproduce an uncertainty in the reddening correction we adopted a normal distribution 
($3\sigma \sim 0.1$ mag). The reddening law was assumed to be $R_V = 3.1$. The 
reddened photometric data of the artificial stars were then obtained from this procedure. 
We corrected for the reddening with the mean value, and then obtained the HRD and age 
distribution as described above. An age spread ($\equiv \tau_{90\%} - \tau_{10\%}$) 
of 0.5 Myr was found for a dispersion of 0.1 mag in reddening.

Another data set of artificial stars were generated using the same procedure as above, but without 
the reddening. We assumed the increase in photometric errors with $V$ magnitude to have 
a similar distribution to the observed one. The errors at a given magnitude were set to have a normal distribution. 
We obtained the age distribution after adding the generated errors to the data. The derived age spread was about 1 Myr. 
A simulation taking into account both the uncertainties in reddening and photometry 
was also carried out. The resultant age spread was also about 1 Myr. Given the results of the 
simulations, photometric errors can be one of the major sources affecting the age spread. Results 
obtained from a single ground-based observation may include that spread. However, there needs to be additional 
sources to explain the observed spread of 5 Myr. \citet{BNLJ05} investigated the influence of 
variability on timescales of a few years, and binarity, on an observed age spread, concluding that 
those factors can contribute only a small fraction of the apparent age spread. Star formation 
history within NGC 1893 turns out to contribute to a spread larger than 0.5 Myr. Details on this 
issue will be discussed in Section 6. Several previous studies have pointed out that evolutionary 
models of PMS stars are likely to overestimate the age of intermediate-mass PMS stars (e.g. 
\citealt{SBL97,SBC04,H99}). However, were the age distribution of these PMS stars shifted toward 
older age, one would expect the age spread to become even wider. We checked the age distribution of 
the PMS members with masses larger than 1.5 $M_{\sun}$. A peak in the distribution appears 
near at 2 Myr. This is slightly older than that found in Fig.~\ref{fig9}. However, the 
fraction of the stars in the mass range is only about 15 per cent of the used sample. The 
contribution of intermediate-mass PMS stars to the observed age spread may not be significant.   

Our age estimate (1.5 Myr) from the MS members is in good agreement with 
that of \citet{PSM11}. \citet{TCER91} and \citet{SPO07} found a somewhat older 
age of 4 Myr from isochrone fitting. Current studies \citep{NMIB07,SAW11} 
have confirmed that the most massive O-type stars (BD +33 1025, HDE 242908, HDE 
242926, and HDE 242935) are on the MS. However, the brightest stars in figure 
11 of \citet{TCER91} are located on the giant loci. It implies that their age is 
overestimated due to the incorrect distance (4.4 kpc). In the case of \citet{SPO07}, the 
age estimation seems to be associated with the limited age range (from 4 Myr to 16 Gyr) 
of the stellar evolutionary models they used. For the PMS stars, our age estimates are 
commensurate with those of previous studies \citep{NMIB07,SPO07,PSM11,PSC13}.

\section{THE INITIAL MASS FUNCTION}

\begin{figure}
\includegraphics[height=0.45\textwidth]{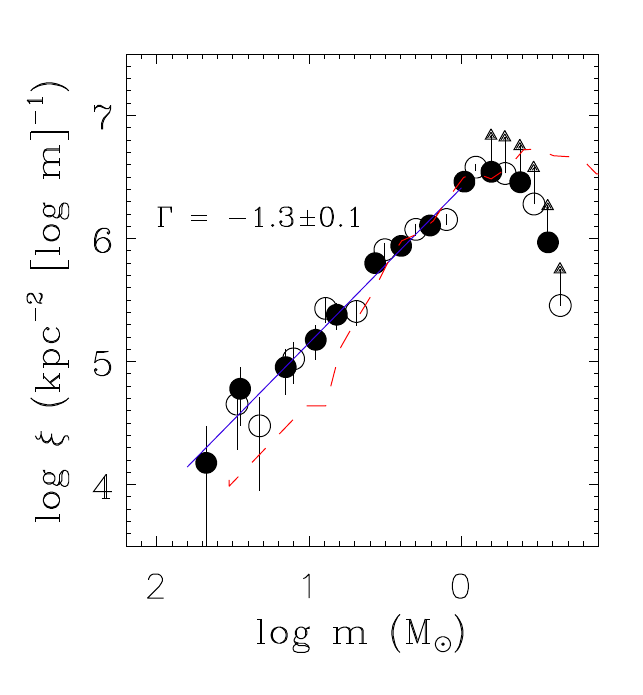}
\caption{The IMF of NGC 1893. To avoid the binning effect we shifted 
the mass bin by 0.1 and rederived the IMF (open circle) using the same procedure. 
The IMF of NGC 2264 (dashed line -- \citealt{SB10} ) is overplotted. Arrows are drawn 
for the IMF below completeness limit. See the main text for details. }
\label{fig10}
\end{figure}

The IMF is an essential tool to understand star formation processes. Therefore a lot of studies on 
IMF, especially on the universality of IMF have been carried out. \citet{BCM10} 
discussed the variation of the IMF with environmental conditions or cosmic time 
based on many previous studies. However, none of the studies could arrive at any firm 
conclusion on its variation with environmental conditions, such as different 
metallicity, the galactic tidal force, suppression of the low-mass star formation 
by a number of high-mass stars, etc. In this context, NGC 1893 is one of ideal 
laboratories to study the IMF of young open clusters in the outer Galaxy. In this work, we can derive reliable 
IMF over a wide mass range, from low-mass PMS stars to massive O-type stars.

The mass of MS members is estimated by comparing their position in the 
HRD to the evolutionary tracks of \citet{EGE12} with various initial masses. For
PMS members, the PMS evolutionary tracks of \citet{SDF00} were used to 
determine the individual mass of each star. The sum of the masses of the members yields a total mass of 
1,300 $M_{\sun}$ for the cluster. This is a lower limit because a large number of sub-solar mass 
PMS stars below the completeness threshold are not taken into account. In order to 
derive the IMF we counted stars within a given mass bin ($\Delta \log m = 0.2$). The star counts were then 
normalized by the logarithmic mass bin and the observed area. The IMF of 
NGC 1893 is shown in Fig.~\ref{fig10}. We shifted the mass bin by 0.1 and 
rederived the IMF (open circle) using the same procedure as above to avoid 
binning effects.

While the IMF of other young open clusters, such as NGC 2244 \citep{PS02}, 
NGC 2264 \citep{PSBK00}, NGC 6530 \citep{SCB00}, and IC 1848 \citep{LSKBK14}, 
shows a significant dip between $3 M_{\sun}$ and $1 M_{\sun}$, that of NGC 1893 
(this work) exhibits a smooth increase down to the completeness limit (see also \citealt{SB10,HSB12,SSB13}). 
It implies that complete membership selection is crucial for the study of the IMF. 
\citet{SB10} presented the IMF of NGC 2264 complete down to 0.25 $M_{\sun}$ using 
the X-ray source catalogue of \citet{FMS06}. The luminosity of a PMS star with 
a mass of 0.2 $M_{\sun}$ at 3 Myr was obtained from the PMS models of \citet{SDF00}, and its 
X-ray luminosity was estimated by assuming the saturation level of X-ray luminosity ($L_{\mathrm{X}}/L_{\mathrm{bol}} 
= -3$). A ratio of the X-ray luminosities of PMS stars in NGC 1893 and NGC 2264 at a completeness limit 
was computed by considering the distance and the total exposure time of each X-ray 
observation. The X-ray luminosity of the lowest-mass PMS star in the complete sample 
of NGC 1893 was estimated to be $0.59\times10^{-3} L_{\sun}$. From the 
dynamo saturation level the luminosity of the star was about $0.59 L_{\sun}$, which could 
be found in the PMS model of \citet{SDF00} for 0.4 $M_{\sun}$ star at 1.2 Myr (the peak age in 
Fig.~\ref{fig9}). Hence, the X-ray source catalogue \citep{CMP12} is complete down to 
$0.4 \ M_{\sun}$. That mass limit is in good agreement with that expected 
from \citet{CMP12}. Hence, the IMF derived in this work is complete down to 
$1 M_{\sun}$ according to the completeness of our photometry. 

The slope ($\Gamma$) of the IMF is about $-1.3 \pm 0.1$ for all the observed 
members with mass larger than $1 M_{\sun}$. It appears to be consistent with the single IMF 
\citep{Sp55} and the Kroupa IMF \citep{K01,K02}. However, the spatial coverage in this 
work is slightly dislocated from that of the {\it Chandra} observation as shown 
in the left-hand side panel of Fig.~\ref{fig1}. We derived the IMF only for the members observed within the 
{\it Chandra} field of view. The slope of this IMF is almost the same as the result above. 
The different spatial coverage may not result in a serious discrepancy. \citet{SPO07} also obtained 
the same slope over the mass range of $0.6 M_{\sun}$ to $17.7 M_{\sun}$ from 
the statistically cleaned ($V,V-I$) CMD. A further steep slope of $\Gamma = -1.6\pm0.3$ 
was derived by \citet{MJD95} for stars with mass larger than $10 M_{\sun}$. Although our research 
for NGC 1893 was made by assuming a solar composition, it is interesting to note that the 
slope of the IMF indicates the standard Salpeter/Kroupa IMF. It implies that the property 
of star formation in the outer Galaxy may be similar to that found in the solar 
neighbourhood.

We also overplotted the IMF of NGC 2264 \citep{SB10} 
in Fig.~\ref{fig10} to compare its shape with that of NGC 1893. The IMF of NGC 1893 is flatter 
than that of NGC 2264 ($\Gamma = -1.7$) for masses greater than $3 M_{\sun}$. However, NGC 1893 hosts 
5 confirmed O-type stars, while there is a binary system consisting of 2 O-type stars 
in NGC 2264 \citep{GMH93}. Furthermore, the binary system of the latter was 
assumed to be a single star due to the limited information on the multiplicity of 
the observed stars. It is difficult to meaningfully compare the IMF of the two 
clusters in the high-mass regime due to the small number of massive stars. 
Hence, we are investigating the integrated features of the IMF derived from 
various SFRs and young open clusters in a homogeneous way based on a 
series of studies  to arrive at a firm conclusion on the universality or diversity of the IMF. 

\begin{figure}
\includegraphics[height=0.45\textwidth]{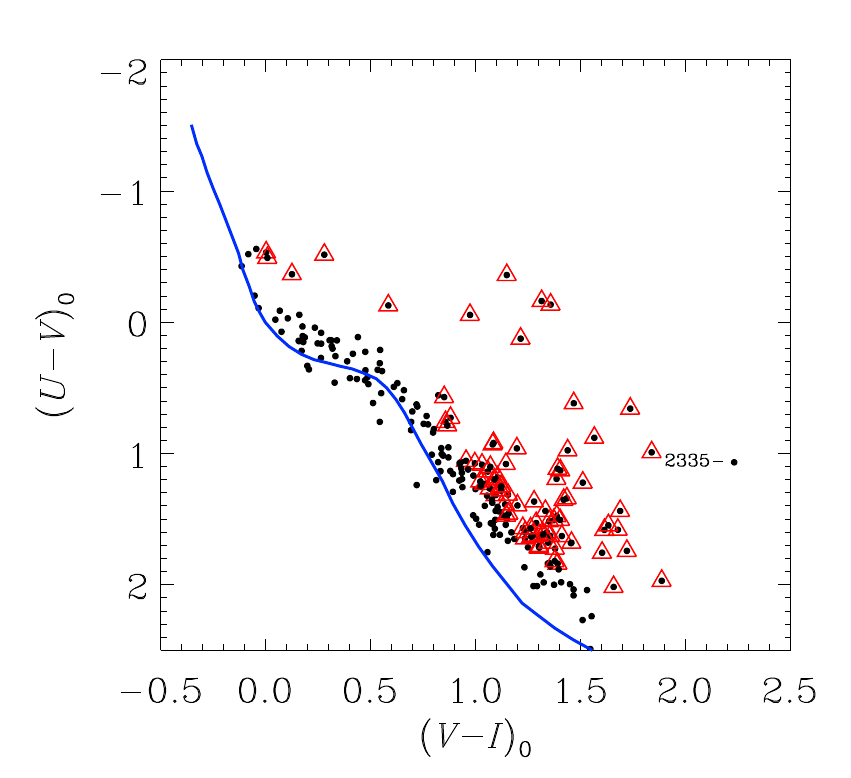}
\caption{The $(U-V)_0$ vs. $(V-I)_0$ diagram. Several PMS stars exhibit 
a strong UV excess arising from accretion activity. A total of 82 UV excess stars (triangles) 
are found. }
\label{fig11}
\end{figure}

\section{MASS ACCRETION RATES OF PMS STARS WITH UV EXCESS }

The mass accretion rate is a useful physical quantity to assess the evolution 
of circumstellar discs and to understand the formation processes of a planetary system around 
PMS stars. Since the application of the magnetospheric accretion 
model to PMS stars \citep{US85,BBB88,K91}, the physical 
quantities associated with the accretion process were studied through observation 
and modelling UV excess emission or/and spectral lines such as H$\alpha$, Pa $\beta$, 
Br$\gamma$, [O {\scriptsize \textsc{I}}] $\lambda$6300, etc, within the paradigm. Many 
studies have been devoted to investigating the mass accretion rates of PMS stars 
in several nearby SFRs, such as the Chamaeleon I, ONC, $\rho$ Ophiuchi, 
and Taurus region, etc (\citealt{FBW09,FKB13,MRR12,MHC98,MHCBH03,MLBHC05,NTM04,NTR06,RHS00}, and 
references therein). The goal from our series of studies is to obtain the 
mass accretion rate of PMS stars ($0.5 M_{\sun} \leq M_{\mathrm{stellar}} \leq 5 M_{\sun}$) 
in many young open clusters within 3 kpc of the Sun in a homogeneous 
manner. We have photometrically estimated the mass accretion rates of PMS 
stars in the young open cluster IC 1848 \citep{LSKBK14} as part of the SOS project. 
Since NGC 1893 is young enough to study the mass accretion rates of intermediate-mass 
PMS stars, this study can provide more information on the $\dot{M} \propto M^{\mathrm{b}}_{\mathrm{stellar}}$ 
relation for such stars together with the results for IC 1848. In this section we estimate the accretion 
luminosities and mass accretion rates of the PMS members with a UV excess using the empirical 
relation derived by \citet{GHBC98} and compare the results with that of other 
studies which used independent ways to estimate accretion luminosities for 
different nearby SFRs.

\begin{figure*}
\includegraphics[height=0.45\textwidth]{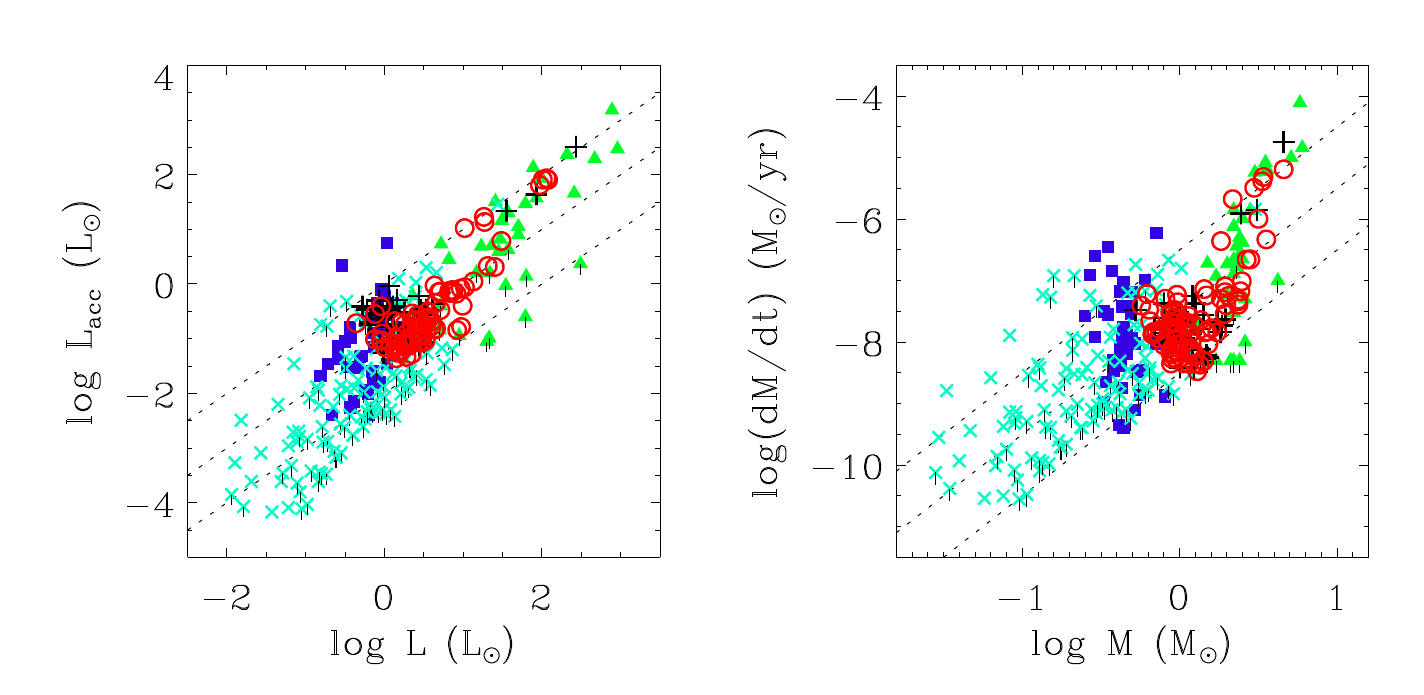}
\caption{Accretion luminosity vs. stellar luminosity (left) and mass accretion rate vs. stellar 
mass (right). Open circles are from this study, while squares, triangles, crosses, and plus signs represent the mass accretion rates 
derived in different SFRs by \citet{HCGD98}, \citet{MCM11}, \citet{NTR06}, and 
\citet{LSKBK14}, respectively. Dotted lines represent the relation $L_{\mathrm{acc}} \propto L_{\mathrm{stellar}}$ 
and $\dot{M} \propto M^2_{\mathrm{stellar}}$ with an arbitrary constant.}
\label{fig12}
\end{figure*}

The PMS members with UV excess emission can be identified in the $[(U-V)_0, (V-I)_0]$ 
diagram as shown in Fig.~\ref{fig11}. These PMS members appear to be bluer in $U-V$
than normal MS stars at a given $(V-I)_0$. Although there may be a considerable scatter, a few stars exhibit a 
prominent UV excess. \citet{RHS00} have discussed several sources affecting the calculated 
UV excess, such as chromospheric activity, accretion activity, 
and the difference between the intrinsic colours of MS and giant stars and examined the 
influence of chromospheric activity on the size of the UV excess using field dMe stars and 
young stars in the Taurus-Auriga SFR. The authors suggested that the limit 
of UV excess from chromospheric activity is about -0.5 mag. This value is adopted as a 
criterion to select those PMS members with UV excess emission as used in \citet{LSKBK14}. A total of 
82 members were identified as PMS stars with a UV excess as shown in Fig.~\ref{fig11} (triangle). 
It is worth noting that the star ID 2335 with small photometric errors has a blue $U-V$ colour. Given its 
position in the HRD ($T_{\mathrm{eff}} = 3.550$ and $M_{\mathrm{bol}} = 1.995$) the star may be a very young PMS star evolving 
along the Hayashi track. Since the luminosity and effective temperature of the star are out of range 
of the evolutionary tracks published by \citet{SDF00}, we could not obtained its mass, thereby excluding 
that in the determination of mass accretion rate as well as the IMF.

We computed the $U_{\mathrm{exp}}$ magnitude expected for a normal photosphere of MS stars and 
the extinction-corrected $U_{0}$ magnitude of stars with a UV excess then transformed 
them to luminosity ($L_{\mathrm{exp}}$ and $L_{U,0}$) using a bandwidth (700 \AA) and zero flux of 
$4.22 \times 10^{-9} \mathrm{erg} \ \mathrm{s} \ \mathrm{cm}^{-2}$ \AA$^{-1}$ for the Bessell $U$ filter \citep{C00}. 
The accretion luminosity $(L_{\mathrm{acc}})$ was estimated from the relation of \citet{GHBC98}: 

\begin{equation}
 \log (L_{\mathrm{acc}}/L_{\sun}) = 1.09 \log (L_{U,\mathrm{exc}}/L_{\sun}) + 0.98
\end{equation}

\noindent where $L_{U,\mathrm{exc}} \equiv L_{U,0} -L_{\mathrm{exp}}$. 

In order to obtain mass accretion rates we estimated the mass of 
individual stars from the evolutionary models of \citet{SDF00} as described in the previous 
section. The radii of stars were obtained by using the effective temperature and 
bolometric magnitude of PMS stars. The mass accretion rate of PMS stars was 
estimated by using the mass ($M_{\mathrm{PMS}}$), radius ($R_{\mathrm{PMS}}$), and accretion 
luminosity ($L_{\mathrm{acc}}$) with the equation below \citep{HCGD98,GHBC98}:

\begin{equation}
\dot{M} = L_{\mathrm{acc}} R_{\mathrm{PMS}} / 0.8 G M_{\mathrm{PMS}} 
\end{equation}

\noindent where $G$ and $\dot{M}$ represent the gravitational constant and 
mass accretion rate, respectively. We present the accretion luminosity and mass 
accretion rate of 82 PMS members in Fig.~\ref{fig12}. Because of the 
large distance to NGC 1893, our study only covers the mass ranges between $0.6 M_{\sun}$ 
and $5 M_{\sun}$. We also plotted the results of other studies for different SFRs 
(\citealt{HCGD98,NTR06,MCM11}, and data therein) which used independent 
ways to estimate the accretion luminosity as well as those of our previous 
study for the young open cluster IC 1848 \citep{LSKBK14}. The accretion luminosity 
and mass accretion rate in this study seem to be compatible with those of other studies. 
The mean mass accretion rate of stars with mass smaller than $2 \ M_{\sun}$ is about $1.6 
\times 10^{-8} M_{\sun} \ \mathrm{yr}^{-1}$. 

\begin{figure*}
\includegraphics[height=1.0\textwidth]{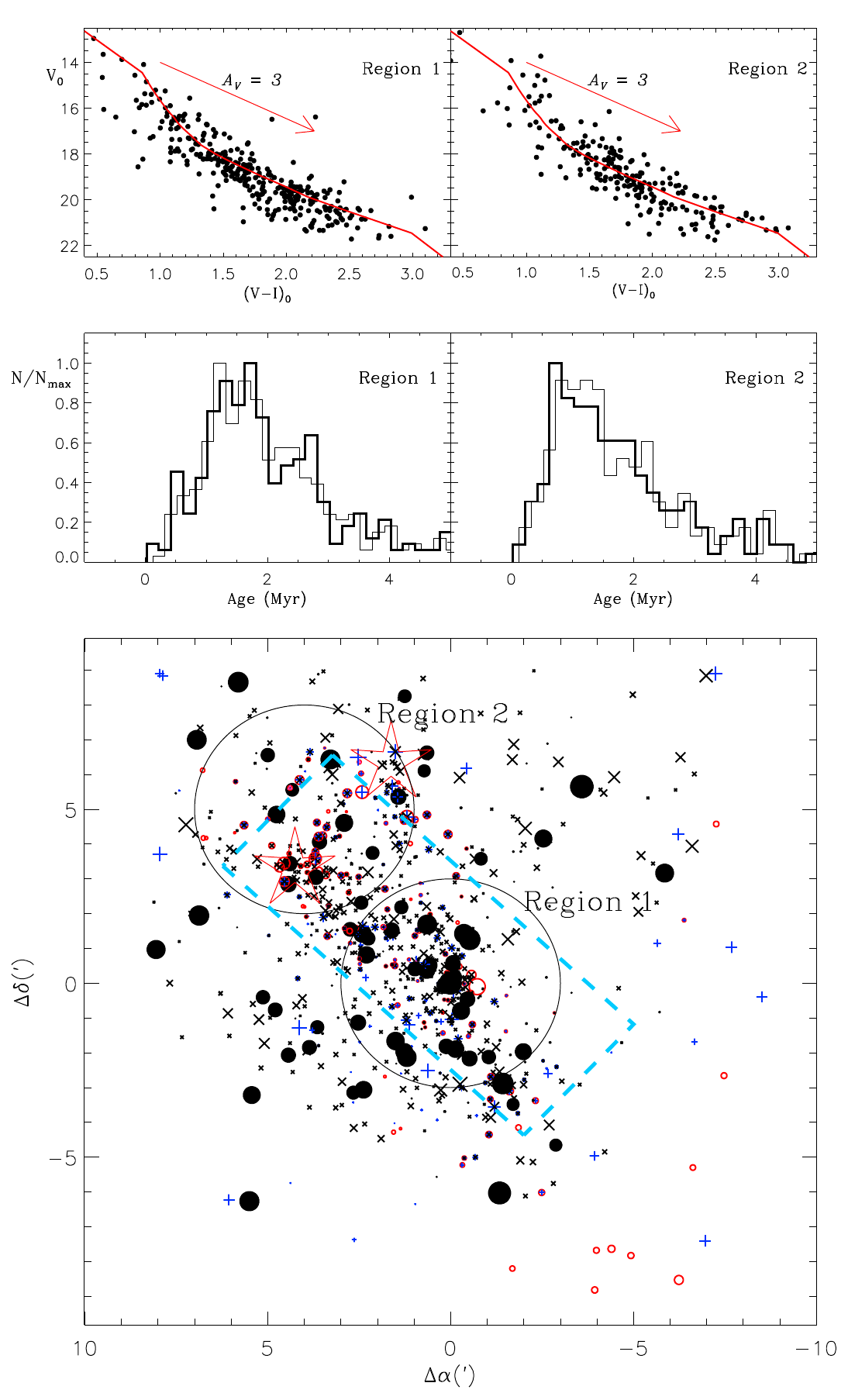}
\caption{Age distribution of the PMS stars in two different regions. The finder chart in the lower panel shows the 
spatial distribution of the members. The star marks denote the location of the heads of the emission nebulae Sim 
129 and 130. The region centred on the cluster centre (Region 1) and that on the vicinity of the emission 
nebulae (Region 2) are outlined by circles. A rectangular box (dashed line) represents a region that was used 
to search for the age variation of PMS stars with the distance from the cluster core. The other symbols are the 
same as Fig~\ref{fig3}. The reddening-corrected CMDs of the PMS members in each region are plotted in 
the upper panels. The solid line (red) represents the 1.5 Myr isochrone from the PMS evolutionary models of 
\citet{SDF00}. The arrow denotes a reddening vector corresponding to $A_V = 3$ mag. The age distributions 
of PMS stars within the regions are shown in the middle panels. The histograms are based on different binning 
of the same stars (thick and thin solid lines). Each distribution is normalized by a peak value ($N_{\mathrm{max}}$). }
\label{fig13}
\end{figure*}

Attempts to understand the accretion disc have been based on the viscous disc model 
\citep{LP74}. Currently, the theory that magnetorotational instability (MRI) in the ionized 
disc plays a very important role in the angular momentum transport has been accepted 
as the major viscosity mechanism. Since MRI operates efficiently in a highly 
ionized accretion disc, there needs to be ionizing sources from the central star or within the 
disc. \citet{GNI97} suggested that stellar X-rays could be a feasible ionizing source of a circumstellar disc. 
\citet{MHCBH03} pointed out that it is difficult to understand how the variation of X-ray 
luminosity with the mass of the central stars affects the $\dot{M}$-$M_{\mathrm{stellar}}$ 
relation, although the correlation between X-ray luminosity and mass was found by many 
other studies \citep{FGGHT03,FDM03,MS02,PZ02}. On the other hand, \citet{HDCM06} 
suggested that the $\dot{M}$-$M_{\mathrm{stellar}}$ relation can be interpreted as a complicated 
mix of different accretion mechanisms, such as layered accretion, fully viscous 
accretion, and gravitational instability.

In the left-hand side panel of Fig.~\ref{fig12}, we plotted the 3 dotted lines corresponding 
to $L_{\mathrm{acc}}/L_{\mathrm{stellar}} = 0.01, 0.10, 1.00$ as shown in \citet{MCM11}. With the estimates 
of other studies the accretion luminosity of the PMS stars ($\log L \leq  1$) is roughly 
consistent with $L_{\mathrm{acc}}/L_{\mathrm{stellar}} = 0.10$, while the accretion luminosity of stars 
with $\log L > 1$ is close to $L_{\mathrm{acc}}/L_{\mathrm{stellar}} = 1$ after an abrupt increase. 
A similar aspect is found in the relation between the mass accretion rate and the mass of the central stars. The mass 
accretion rate of PMS stars with mass larger than $2.5 M_{\sun}$ is approximately 10 times 
higher than that expected from the $\dot{M}$-$M_{\mathrm{stellar}}$ relation for lower mass stars. 
\citet{MCM11} found that the age of Herbig Ae/Be stars in their sample was systematically 
younger than that of low-mass stars, and the steep slope in the mass accretion rate with 
respect to the stellar mass was interpreted as a consequence of the huge accretion of 
such stars in the early stages. However, \citet{GNTH06} obtained lower mass accretion rates for the same 
Herbig Ae stars. The estimates of \citet{DB11} (except the upper limits), also reveal 3 times lower values for 
Herbig Ae/Be stars. Practical uncertainties in calibration schemes, reddening correction, and 
distance can result in systematic differences in the mass accretion rates for the Herbig 
Ae/Be stars between authors. 

While X-ray emission from the central stars plays a crucial role in the operation of MRI within 
the accretion discs of low-mass PMS stars, the role of X-ray emission in the accretion processes of intermediate-mass 
PMS stars is still uncertain (\citealt{HDCM06} and references therein). In our samples 4 out of 6 
intermediate-mass PMS stars with UV excess emission ($> 2.5 \ M_{\sun}$) have low X-ray 
luminosities of $ L_{\mathrm{X}} \sim 1.6 \times 10^{30}$ erg s$^{-1}$, that are comparable 
to those of low-mass counterparts ($\sim 1 \ M_{\sun}$). It implies that X-ray emission from the 
central star may be insufficient to operate MRI. On the other hand, gravitational instability is one of 
the possible mechanisms of angular momentum transfer within circumstellar discs. According 
to \citet{HDCM06} the disc mass of PMS stars derived from dust emission 
(e.g. \citealt{AW05}) is an order of magnitude lower than that required to operate gravitational 
instability in the disc. They suggested that the smaller derived disk masses may result from an 
inappropriate accounting for the opacity variation due to dust growth and that the formation of planets 
within the disk and the spiral structure of the disk are indicating larger disk masses
that would enable gravitational instability to operate. In this context, the structure and geometry 
of accretion discs may be an essential issue in understanding the evolution of the discs in the 
intermediate-mass PMS star systems. However, the details are not within our scope of study 
based on such limited data. Those PMS stars with a UV excess are good candidates for 
spectroscopic follow-up observations in the future.

\section{TRIGGERED STAR FORMATION IN NGC 1893}

The two emission nebulae Sim 129 and 130 attract many astronomers to study triggered 
star formation mechanisms in the cluster (e.g. \citealt{MSBPB07,NMIB07,SPO07,PSC13}). This 
is because the glowing heads of the nebulae face the centre of the cluster. A number of H$\alpha$ 
emission stars were identified in the vicinity of the nebulae \citep{MN02,MSBPB07,NMIB07,SPO07}. 
There have been a few attempts to investigate the age sequence of the PMS stars from the centre of 
cluster to the emission nebulae (e. g. \citealt{SPO07,PSC13}). However, the radial variation in 
the ages seems to be rather ambiguous (see the right-hand side panel in figure 7 of 
\citealt{PSC13}). In order to investigate convincing evidence for the triggered star formation 
we have examined the age distribution of the PMS stars with distance from the 
centre of the cluster.  

As shown in Fig.~\ref{fig13} two different regions were selected in this analysis. One region (Region 1) was 
centred on the O6.5V((f))z star HDE 242935 \citep{SAW11}, and the other region (Region 2) 
was located in the vicinity of the two emission nebulae. The diameter of the regions is about 
6 arcmin. The upper panels in Fig.~\ref{fig13} exhibit the reddening-corrected CMDs of the 
PMS members in the regions. The relative number of stars brighter than the 1.5 Myr isochrone 
(solid line) at a given colour appears to be larger in the Region 2. The reddening values 
of the stars in the regions were estimated by using the same relation as shown in 
the middle panel of Fig.~\ref{fig5}, being within a similar range [$E(B-V) = 0.42$ -- 
0.64 mag] although the differential reddening in the Region 2 is less significant 
than that in the Region 1. Hence, the luminous PMS members in the Region 2 are not 
related to any systematic errors in reddening correction. Given that the reddening vector (arrow) is almost 
parallel to the slope of the PMS isochrone, the difference between the stars in the Region 1 
and 2 may be associated with their age difference rather than the variation of reddening.   

\begin{figure}
\includegraphics[height=0.45\textwidth]{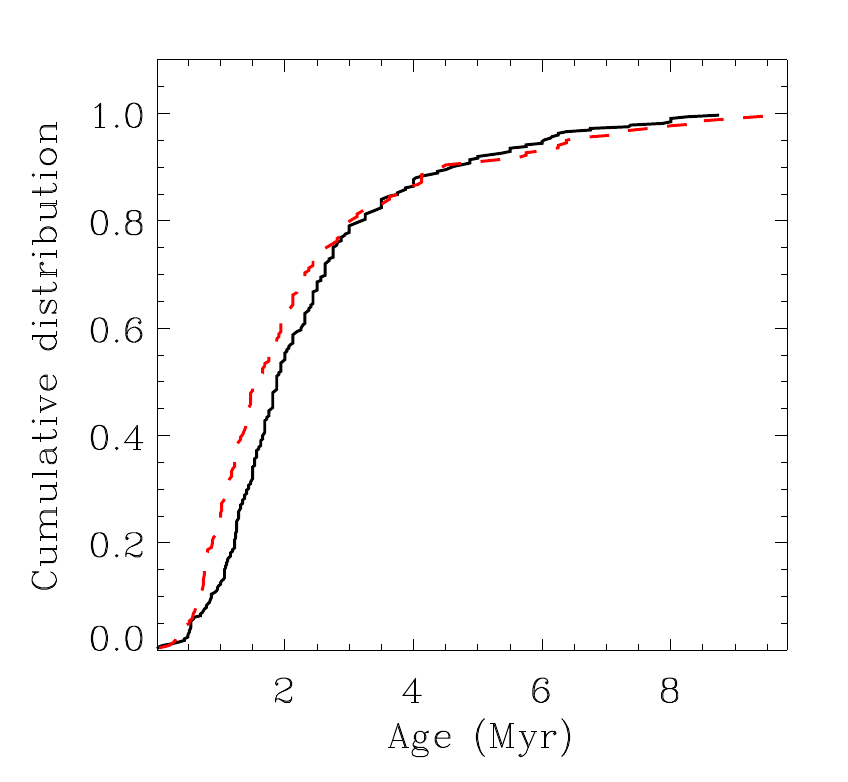}
\caption{The cumulative age distribution of the stars in Region 1 and 2. The solid line and 
dashed line correspond to Region 1 and 2, respectively. See the main text for details. }
\label{fig14}
\end{figure}

We counted the number of stars within a given age bin, where the bin size was about 
0.2 Myr. The age distributions of the PMS members in the Region 1 and 2 are shown in the middle panels, respectively. 
The histograms are normalized by a peak value of each distribution. To avoid a 
binning effect another age distribution (thin solid line) was obtained by shifting the age bin by 0.1 Myr. 
The peak of the age distribution in Region 1 appears at 1.5 Myr, while that in Region 
2 is slightly shifted toward a younger age ($\sim$ 1 Myr). The age difference between the two groups 
is about 0.5 Myr. The PMS stars close to Sim 129 and 130 are likely to be younger than those 
around the centre of the cluster (see also \citealt{PSC13}). We present the cumulative 
age distribution of the PMS stars in each region in Fig.~\ref{fig14}. The figure shows that the 
stars located in Region 2 are slightly younger than those in Region 1. According to the 
Kolmogorov-Smirnov test, the possibility that the PMS stars in the two regions have the 
same origin is about 0.1 per cent. A statistical approach also supports the age difference between the 
two regions. 

\begin{figure}
\includegraphics[height=0.33\textwidth]{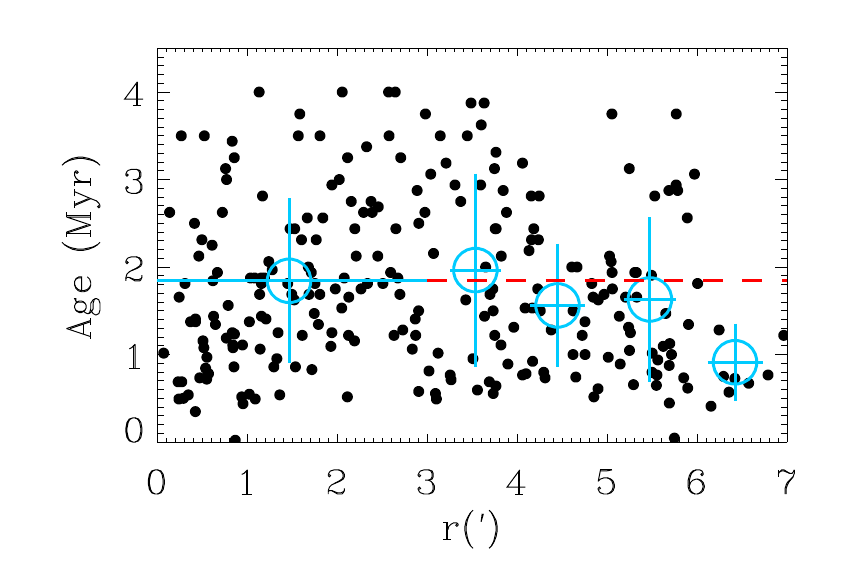}
\caption{The sequential age variation of PMS stars from the O-type star HDE 242935 toward the 
emission nebulae Sim 129 and 130. The mean age at a given distance bin is shown as a circle 
with an error bar which is equivalent to a standard deviation at the same bin. The dashed line 
represents the mean age of the stars within the cluster core ($\sim$ 3 arcmin) }
\label{fig15}
\end{figure}

We investigated whether or not the formation of the younger age group was triggered by the high-mass stars. 
If the O-type stars in the centre of NGC 1893 are responsible for the formation of the new generation 
stars, one may expect the age of the PMS stars to be sequentially younger with distance from 
the cluster centre. We selected the PMS members within a region encompassing the cluster core 
and periphery of the two emission nebulae (the rectangular box in Fig.~\ref{fig13}). The mass of 
the PMS members was limited to 0.5 -- 1.5 $M_{\sun}$ because of the fact that current
evolutionary models tend to overestimate the age of intermediate-mass PMS stars 
\citep{Hd97,SBL97,SBC04,H99}. We only considered the PMS members younger than 4 Myr 
to seek a clear age variation. Fig.~\ref{fig15} shows the age variation of PMS stars within the 
rectangular box in Fig.~\ref{fig13} to be away from the cluster core. Since new generation 
stars are expected to have formed just after the formation of the cluster core the first 
bin includes all the stars in the core. The projected extent is about 3 arcmin which was inferred 
from the surface density profile. The mean age (circle) reasonably attenuates toward the emission nebulae 
out of the cluster core ($> 3$ arcmin) although there is a rather large scatter arising from 
many sources of random errors, such as photometric errors, incomplete reddening correction, excess emissions from 
both circumstellar discs and accretion activities, and variability (see Section 3.4 and the review of \citealt{SHJ13} 
for details). On the other hand, the projected distance between the ionizing source HDE 242935 and the 
emission nebula Sim 129 is about 5.6 arcmin ($\sim$ 5.6 pc). If we assume a sound 
speed of 9.6 km/s in an expanding H {\scriptsize \textsc{II}} region ($T \sim 8,000 \mathrm{K}$), the 
travel time of the materials being compressed from the the O-type star to the current position 
of the emission nebula is about $5.7 \times 10^5$ yr. This is in reasonably agreement with the 
age difference between Region 1 and 2. Our results well support the argument 
of previous studies \citep{SPO07,PSC13} that the formation of the young stars in the vicinity 
of Sim 129 and 130 may have been triggered by the source HDE 242935 in the centre of the cluster.

There are two convincing mechanisms for triggered star formation. The first is the `collect and 
collapse' (hereafter CC) process proposed by \citet{EL77}. Massive stars create an H {\scriptsize \textsc{II}} region 
in their natal cloud. Subsequently, the expansion of the H {\scriptsize \textsc{II}} region contacts
the molecular cloud and generates a shock front at the boundary. Gas and dust in the cloud are 
swept away and compressed into small clumps as the shock front is propagated toward the outside of the 
cloud. In the end, the clumps become gravitationally unstable at a critical density, and so star formation 
takes place within the clumps. For instance, the new generation of stars along the border of the giant 
H {\scriptsize \textsc{II}} region W5 are thought to have been formed by the CC mechanism \citep{KAG08}. 
The other mechanism is the `radiatively driven implosion' (hereafter RDI) process, in which strong 
UV radiation from massive stars leads pre-existing density-enhanced clumps to 
gravitationally collapse. The typical observational evidence are the presence of cometary 
globules, bright-rimmed molecular clouds, and the spatial distribution and age variation of 
PMS stars (e. g. \citealt{LL94,LLC97,TWM04,SPO07,KAG08,PSC13}). For the 
formation of the younger PMS population within NGC 1893, \citet{PSC13} advocates a scenario 
that the formation of new generation stars has been controlled by the RDI process \citep{SPO07,MSBPB07}. 

\citet{SPO07} investigated the structure of the surrounding interstellar medium using the 
{\it Midcourse Space Experiment} $A$-band intensity and the {\it NRAO VLA Sky 
Survey} radio contour maps. Few emission clumps were found in the periphery of the cluster, while an 
extended arc-like cloud was located at the western part (see also figure 1 of \citealt{CMP08}). The 
entire structure may have been created by the expansion of the H {\scriptsize \textsc{II}} region 
during the early phase of cluster formation. There are two considerable clumps associated with {\it IRAS} sources and an O-type 
star in the extended cloud. These were interpreted as objects escaping from the natal cloud 
\citep{SPO07}. Several PMS stars with either H$\alpha$ or MIR excess emission are sparsely populated in 
the direction of the arc-like cloud (see Fig.~\ref{fig13}). If the current location of the stars is 
related to their birthplace, the expansion of the H {\scriptsize \textsc{II}} bubble may be responsible 
for their formation, i.e. the CC process. However, the connection between the stars and the 
H {\scriptsize \textsc{II}} bubble seems to be uncertain. Other clumps are associated with the prominent emission 
nebulae Sim 129 and 130. The morphology of the nebulae facing toward the cluster core, especially the 
bright rim, is likely indicative of ongoing interaction with the UV radiation emitted from the ionizing 
sources \citep{LL94}. The stellar density in the direction of the nebulae is higher than that in other regions outside 
of the cluster core (see also figure 6 of \citealt{SPO07}). If the distribution of the stars reflects 
the sub-structure of the natal cloud prior to their formation \citep{EEPZ00}, the density of the 
cloud might be partly high. This assumption seems to explain the presence of pre-existing 
density-enhanced clumps. In addition, the ages of the PMS stars near the emission nebulae 
(in Region 2) are similar to the formation timescale of cometary globules ($\sim 1.3$ Myr, 
\citealt{LL94}). Hence, these aspects indicate that the RDI process seems to be a more plausible 
triggering mechanism in the north-east of NGC 1893 rather than the CC process. 

\section{SUMMARY}
NGC 1893 which hosts a large number of young stars, from low-mass PMS stars to massive O-type stars, 
is one of the well-known young open clusters in the Galaxy. The cluster gives us a chance to study 
star-forming activities in the outer Galaxy, which is (putatively) a different environment 
from the solar neighbourhood \citep{CMP08} although its chemical composition is somewhat uncertain 
\citep{RBDF93,DC04}. We carried out $UBVI$ and H$\alpha$ photometry for the cluster as part of the SOS project. This 
study provided not only homogeneous photometric data but also detailed and comprehensive 
results for NGC 1893 as shown below.

Using the photometric properties of early-type stars, 65 early-type (O -- B) 
and 6 A-type MS stars were selected as members. We identified 213 young stars with MIR excess emission from 
{\it Spitzer} IRAC data \citep{CMP08}. From our H$\alpha$ photometry, 126 H$\alpha$ 
emission stars and candidates were also identified, most of which are PMS members. 
In addition, 703 PMS members (including 55 candidates) were found using the X-ray 
source list of \citet{CMP12}. A total of 906 stars were thus selected as members of NGC 1893. 

From the $(U-B, B-V)$ diagram we obtained the reddening of individual 
early-type stars yielding a mean reddening of $\langle E(B-V) \rangle = 0.56 \pm 0.08$ mag. 
As seen in most young star clusters or associations, differential reddening is not 
negligible. We constructed the reddening map from the 
differential surface density map between the $V$ and 3.6 $\mu m$ bands. The 
photometric data from the optical to MIR for the early-type stars in the observed 
regions allowed us to test the reddening law of NGC 1893. We confirmed the normal 
reddening law ($R_V = 3.1$) toward the cluster from various colour excess ratios over 
a wide wavelength coverage from the optical to the MIR. 

A careful ZAMS fitting to the lower 
boundary of the MS band was carried out in the reddening-independent 
$Q_{V\lambda}$-$Q^{\prime}$ planes, and we obtained a distance modulus of 
$12.7 \pm 0.2$ mag, which is in reasonably good agreement with that of previous $UBV$ 
photometric studies. The isochrone fitting to the most massive stars in the HRD gives an age of 1.5 Myr, 
and the age distribution of PMS stars shows the median age of 1.9 Myr with an age spread of 
5 Myr. The age scale between the MS and PMS members appears to be well consistent. 
The age spread of the PMS members is very similar to that found 
in many other young open clusters.

Based on a careful membership selection, the IMF complete down to $1 M_{\sun}$ 
was derived from cluster members. The slope ($\Gamma = -1.3$) is consistent with the 
Salpeter/Kroupa IMF. It implies that the property of star formation in the outer 
Galaxy may not be far different from that found in the solar neighbourhood. 
This will be one of the valuable results to help arrive at a firm conclusion 
on the universality of the stellar IMF in the future.

We found 82 PMS stars with strong UV excess emission from our photometry. The $U$ luminosity 
was transformed to the accretion luminosity by adopting the empirical relation 
from the literature. Finally, the mass accretion rate of the PMS stars was estimated by 
using the accretion luminosity, the computed stellar radius, and the stellar mass 
inferred from the PMS evolutionary models. The stars with mass smaller than 2 $M_{\sun}$ 
among the UV excess stars exhibited a mean mass accretion rate of $1.6 \times 10^{-8} 
M_{\sun} \ \mathrm{yr} ^{-1}$. Our result is well consistent with the estimates of other studies. 
We also found that high-mass stars show a higher mass accretion rate than the accretion 
rates expected from the correlation as shown in previous studies \citep{MCM11,LSKBK14}

The two emission nebulae Sim 129 and 130 are interesting objects with which to study triggered 
star formation processes in NGC 1893. As done in previous studies \citep{SPO07,PSC13}, we 
investigated the age sequence from the O-type star HDE 242935 to the emission nebulae. 
The age distribution of the PMS members showed a continuously varying age sequence, 
which appears to be more significant than the result of previous studies. This has been interpreted 
as evidence of triggered star formation. On the basis of several discussions, we concluded 
that the RDI process may be the dominant triggering mechanism for new 
generation stars in NGC 1893.

\section*{acknowledgments}
The authors thank the anonymous referee for many useful comments. This work was 
supported by a National Research Foundation of Korean (NRF) 
grant funded by the Korea Government (MEST) (Grant No.20130005318).

%%%%%%%%%%%%%%%%%%%%%%%%%%%%%%%%%%%%

\end{document}